\documentclass[leqno]{amsart}
\usepackage{amsmath,amsthm,amsfonts}
\usepackage{latexsym}
\usepackage{amssymb}
\usepackage[dvips]{epsfig}

\renewcommand{\thesection}{\arabic{section}}

\newtheorem{theorem}{Theorem}[section]
\newtheorem{lemma}[theorem]{Lemma}
\newtheorem{prop}[theorem]{Proposition}
\newtheorem{defi}[theorem]{Definition}

\newtheorem{corollary}[theorem]{Corollary}
\theoremstyle{remark}

\renewcommand{\theequation}{\thesection .\arabic{equation}}
\let\subs\subsection
\renewcommand\subsection{\setcounter{equation}{0}
\gdef\theequation{\thesubsection \arabic{equation}}\subs}
\let\sect\section
\renewcommand\section{\setcounter{equation}{0}
\gdef\theequation{\thesection .\arabic{equation}}\sect}

\newcommand{\cA}{{\mathcal{A}}}
\newcommand{\cB}{{\mathcal{B}}}
\newcommand{\cD}{{\mathcal{D}}}
\newcommand{\cE}{{\mathcal{E}}}
\newcommand{\cG}{{\mathcal{G}}}

\newcommand{\cJ}{{\mathcal{J}}}
\newcommand{\cN}{{\mathcal{N}}}

\newcommand{\cW}{{\mathcal{W}}}

\newcommand{\cT}{{\mathcal{T}}}
\newcommand{\cS}{{\mathcal{S}}}
\newcommand{\cF}{{\mathcal{F}}}

\newcommand{\cL}{{\mathcal{L}}}
\newcommand{\cP}{{\mathcal{P}}}

\newcommand{\IC}{{\mathbb{C}}}

\newcommand{\IR}{{\mathbb{R}}}
\newcommand{\TT}{{\mathbb{T}}}

\newcommand{\tor}{\TT}
\newcommand{\ZZ}{{\mathbb{Z}}}
\newcommand{\IZ}{{\mathbb{Z}}}

\newcommand{\ur}{{\underline{r}}}

\newcommand{\uw}{{\underline{w}}}

\newcommand{\ann}{{\mathcal A}}
\newcommand{\bmo}{{\rm BMO}}

\newcommand{\be}{\begin{eqnarray}}
\newcommand{\ee}{\end{eqnarray}}

\newcommand{\BMO}{{\rm{BMO}}}

\newcommand{\degg}{\mathop{\rm{deg}}}
\newcommand{\dist}{\mathop{\rm{dist}}}
\newcommand{\disc}{\mathop{\rm{disc}}}

\newcommand{\mes}{\mathop{\rm{mes}\, }}
\newcommand{\compl}{\mathop{\rm{compl}}}
\renewcommand{\mod}{{\rm{mod}\, }}
\newcommand{\rsp}{\mathop{\rm{sp}\, }}

\newcommand{\tr}{\mathop{\rm{tr}}}

\newcommand{\car}{\mathop{\rm{Car}}\nolimits}
\newcommand{\imm}{\mathop{\rm{Im}}}

\newcommand{\const}{\mathop{\rm{const}}}

\newcommand{\capo}{\cA_{\rho_0}}

\newcommand{\vep}{{\varepsilon}}

\newcommand{\doppelint}{{\displaystyle{\int\!\!\!\!\int}}}

\newcommand{\strich}{-\!\!\!\!-\!\!\!\!-}
\newcommand{\niint}{{{\strich\!\!\!\!\!\!\!\!\!\doppelint}}}
\newcommand{\strichint}{\mathop{\niint}}
\newcommand{\nn}{\nonumber}

\newcommand{\notint}{{-\!\!\!\!\!\!\int}}
\newcommand{\nint}{\mathop{\notint}}
\newcommand{\la}{\langle}
\newcommand{\ra}{\rangle}
\newcommand{\w}{\omega}
\newcommand{\xwe}{(x,\w, E)}
\newcommand{\xw}{(x, \w)}
\newcommand{\we}{(\w, E)}
\def\beeq{\begin{equation}}
\def\eneq{\end{equation}}
\def\eps{\varepsilon}
\def\les{\lesssim}

\def\cZ{{\mathcal Z}}
\def\bm{\left[\begin{matrix}}
\def\endm{\end{matrix}\right]}
\def\cS{{\mathcal S}}
\def\bad{\cB}
\def\Car{{\rm Car}}
\def\xnull{{x_0}}
\def\Compl{\IC}
\def\half{\frac12}
\def\mape{\bigl(e(x),\omega, E+i\eta\bigr)}
\def\mapen{\bigl(e(x), \omega\bigr)- E-i\eta}
\def\ulm{{\underline{m}}}
\begin{document}
\title{On Schr\"odinger operators with dynamically defined potentials}

\author{Michael Goldstein and Wilhelm Schlag}

\address{Dept.\ of Mathematics, University of Toronto, Toronto, Ontario, Canada M5S 1A1}

\email{gold@math.toronto.edu}

\address{253-37 Caltech, Pasadena, CA 91125, U.S.A.}

\email{schlag@math.princeton.edu}

\thanks{The first author was partially supported by an NSERC grant. The second author was
partially supported by the NSF, DMS-0300081, and a Sloan fellowship.
The authors dedicate this article to Yakov Grigorievich Sinai on the
occasion of his 70th birthday.}

\date{}

\maketitle

\section{Introduction}

The purpose of this article is to review some of the recent work
on the operator
\begin{equation}
(H_x\psi)_n = -\psi_{n-1} - \psi_{n+1} +  \lambda V(T^n x)\psi_n
\label{eq:1}
\end{equation} on
$\ell^2(\ZZ)$, where $T:X\to X$ is an ergodic transformation on
$(X,\nu)$ and $V$ is a real-valued function. $\lambda$ is a real
parameter called coupling constant. Typically,
$X=\tor^d=(\IR/\ZZ)^d$ with Lebesgue measure, and $V$ will be a
trigonometric polynomial or analytic. We shall focus on the papers
\cite{GS1} and~\cite{GS2} by the authors, as well as other work
which was obtained jointly with Jean Bourgain. Our goal is to
explain some of the methods and results from these references. Some
of the material in this paper has not appeared elsewhere in
print\footnote{It is not our intention to provide a systematic and
detailed review of the many developments that have taken place
during the last five to ten years in this vast area and we apologize
to those authors who have been involved in the study of these models
but are not mentioned here.}.

Even more specifically, we will be mostly concerned with the
distribution of the eigenvalues of~\eqref{eq:1}, both on finite
intervals $[-N,N]$ as well as in the limit $N\to\infty$ (and not so
much with Anderson localization). In more technical terms, we are
referring here to the {\em integrated density of states} or IDS. It
is a nondecreasing, deterministic function $N(E)$, and it is related
to the Lyapunov exponent $L(E)$ by means of the Thouless formula
\begin{equation}
\label{eq:thou}
 L(E) = \int \log|E-E'|\, N(dE')
\end{equation}
As usual, we set
\[ L(E) = \lim_{n\to\infty} \frac{1}{n} \int_{X}
\log\|M_n(x,E)\|\,\nu(dx) \] where $M_n$ are the transfer matrices
\[ M_n(x,E) = \prod_{k=n}^1 \bm  V(T^k x)-E & -1 \\
1 & 0 \endm
\]
of \eqref{eq:1}, i.e., the column vectors of $M_n$ are a
fundamental system of the equation $H_x \psi = E\psi$.

One of the  most basic problems related to \eqref{eq:1} concerns the
positivity of the Lyapunov exponent $L(E)$.  More specifically,
there are dynamical systems for which $L(E)$ exhibits a ``phase
transition'' from the region $L(E) > 0$ to that where $L(E) = 0$
when $\lambda$ varies, and there are systems for which $L(E) > 0$
for all values of $\lambda\ne0$.  For instance, one expects that for
the skew-shift $T: \tor^2 \to \tor^2$, $T(x,y) = (x + \w, y+x)$, the
Lyapunov exponent is positive for all $\lambda\ne0$. A rigorous
description of phase transitions for $L(E)$ or the proof of the
absence of such transitions is a primary objective in the study of
\eqref{eq:1}.

In the study of~\eqref{eq:1} much attention has traditionally been
given to the fine properties of the distribution of the eigenvalues
of~\eqref{eq:1}, i.e., the IDS. As observed by Avron and
Simon~\cite{AS}, and by Craig and Simon~\cite{CS} the Thouless
formula implies that the IDS is log-H\"older continuous.
In~\cite{GS1} it was shown that for positive Lyapunov exponents the
IDS is H\"older continuous, and their argument was refined
in~\cite{Bou2} to show that the H\"older exponent remains bounded
below by a positive constant when the Lyapunov exponent approaches
zero. For the almost Mathieu potential $V(\theta)=\cos(\theta)$,
Sinai~\cite{Si1} and Bourgain~\cite{Bou1} observed  perturbatively
(i.e., when $\lambda$ is large), that the IDS is H\"older
$\frac12-\epsilon$ continuous for any $\epsilon>0$. Moreover, it
follows from Sinai's work that this is optimal (see also
Puig~\cite{Pui2}). In the recent paper~\cite{GS2}, the authors
addressed the problem of determining the exact H\"older exponent
non-perturbatively assuming only that $L(E)>0$. Furthermore,
\cite{GS2} treats general potentials (trigonometric polynomials or
analytic functions) and it is shown that the IDS is absolutely
continuous.

An interesting open problem is to understand the behavior of the IDS
under the phase transition \[L(E)
> 0 \longrightarrow L(E) = 0 \]
Recall that for the almost Mathieu case it is known due to the work
by Last~\cite{Last} and by Gordon, Jitomirskaya, Last, and
Simon~\cite{GJLS}, that the Lebesque measure of the spectrum
decreases to zero when the coupling constant approaches the critical
value $\lambda = 2$.

We now set out to describe some of the basic tools developed in the
references \cite{BG}, \cite{GS1}, \cite{BGS}, \cite{GS2}.

\section{Large deviation theorems}

It was shown by F\"urstenberg
and Kesten~\cite{FurKes} that
\[
\lim_{n\to\infty} \frac{1}{n} \log\|M_n(x,E)\| = L(E)
\] for a.e.~$x\in X$. To quantify this
convergence, set
\[ L_n(E) = \frac{1}{n} \int_{X}
\log\|M_n(x,E)\|\,\nu(dx).\]
For certain types of dynamics and $V$ the following {\em large
deviation theorems} (or LDTs) are known to hold for some choice of
$0<\sigma,\tau<1$:
\begin{equation}
\label{eq:LDT1} \nu\big(\{x\in X\::\:| \log\|M_n(x,E)\| - nL_n(E)|
> n^{1-\sigma}\}\big) \le C\exp(-n^\tau).
\end{equation}

For definiteness, let $X=\tor$, $d\nu=dx$ the Lebesgue measure,  and
$Tx=x+\omega \;\mod\ZZ$ with an irrational $\omega$. Moreover, we
shall assume that $V:\tor\to \IR$ is analytic and nonconstant. The
LDTs are known to hold in this case.

To motivate~\eqref{eq:LDT1}, consider first a commutative model case, namely
\[ u(x) = \sum_{k=1}^q \log|e(x)-e(k\omega)| \]
with $\omega=\frac{p}{q}$ and $e(x) = e^{2\pi ix}$. Then
$u(x)=\log|e(xq)-1|$ and $\int_\tor u(x)\, dx=0$  so that for
$\lambda<0$
\begin{equation}
\label{eq:model} \mes\big(\{x\in\tor\::\: u(e(x))< \lambda\})
=\mes\big(\{x\in\tor\::\: |e(x)-1|< e^\lambda\})
\end{equation}
which is of size $e^\lambda$ (here $\mes$ stands for Lebesgue measure). In this model case, $u(x+1/q)=u(x)$.

Returning to $u(x)=\log\|M_n(x,E)\|$, this exact invariance needs
to be replaced by the almost invariance
\begin{equation}\label{eq:ainv}
\sup_{x\in\tor} |u(x)-u(x+k\omega)|\le Ck \text{\ \ for any\ \ }k\ge1.
\end{equation}
The logarithm in our model case is a reasonable choice because of
Riesz's representation theorem for subharmonic\footnote{This is defined
to mean upper semicontinuous and satisfying the sub-mean value property.} functions (see
Levin~\cite{levin}) applied to the function $u(z)=
\log\|M_n(z,E)\|$:

\medskip {\em Let $u(z)$ be a subharmonic function on some domain
$\Omega\subset\IC$. Then there exist a positive measure $\mu$
(called the Riesz measure), finite on all compact sub-domains
$\Omega'\subset \Omega$ so that
\begin{equation} \label{eq:riesz}
 u(z) = \int_{\Omega'} \log|z-\zeta|\,
\mu(d\zeta) + h(z) \quad \forall z\in\Omega'
\end{equation}
where $h$ is harmonic on $\Omega'$. }

If $u\in C^2(\Omega)$ then subharmonicity is the same as $\Delta u\ge0$ and  $\mu=\Delta u$.
Moreover, in this case \eqref{eq:riesz} is an instance of Green's formula. The general
case follows by taking limits. In what follows, we will also make use of the following
bounds on $\mu$ and $h$: If \eqref{eq:riesz} holds and $|u|\le K$ on $\Omega'$,
then
\begin{equation}
\label{eq:mu_bd}
 \mu(\Omega'')+\|h\|_{L^\infty(\Omega'')}\les K
\end{equation}
where $\Omega''\subset\Omega'$ (as a compact sub-domain).

\subsection{Cartan estimates}

One way of deriving~\eqref{eq:LDT1} from \eqref{eq:ainv}
and~\eqref{eq:riesz} is via Cartan's estimate for subharmonic
functions:

\medskip{\em Given a finite positive measure $\mu$ on $\IC$
and $H>0$ there is a (possibly infinite) collection of disks
$\{D(z_j,r_j)\}$ so that\footnote{Exact numerical values of the
constants $C$ are  known, see Levin. Also, one can replace $\sum r_j$ with $\sum r_j^\eps$
for any $\eps>0$, which implies that $\{u=-\infty\}$ has
Hausdorff dimension zero.}
\[ \sum_j r_j < CH \text{\ \ and\ \ } u(z) > -C\|\mu\| \log H
\quad \forall z\in\IC\setminus \bigcup_j D(z_j,r_j)\] }

\medskip The most obvious example here is of course $\mu =
n\delta_0$. More generally, for the measures $\mu=\sum_k
\delta_{z_k}$ with some finite collection $\{z_k\}_k\subset\IC$ of (not necessarily distinct)
points, Cartan's theorem becomes a statement about polynomials
which already captures all the main features. In order to
prove~\eqref{eq:LDT1} we will impose the Diophantine condition
\begin{equation}
\label{eq:strong_dio}
 \|n\omega\|\ge \frac{c}{n^a} \quad\forall n\ge 1
\end{equation}
where $c=c(\omega)>0$ and\footnote{The upper bound of 2 here is more of a cosmetic nature}
 $2>a>1$. Here $\|\cdot\|$ measures the
distance to the nearest integer.

\begin{proof}[Proof of \eqref{eq:LDT1}]
Write
\[ u(z)=\log\|M_n(z,E)\| = \int \log|z-\zeta|\, \mu(d\zeta) + h(z) \]
on some open rectangle $R$ which contains\footnote{recall that
$u(x)=\log\|M_n(x,E)\|$ is a one-periodic function} $[0,1]$. Then $0\le u(z)\les n$ and thus
$\mu(R')\les n$ where $R'\subset R$ is a slightly smaller
rectangle, as well as $\|h\|_{L^\infty(R')}\les n$, see~\eqref{eq:mu_bd}. Fix a small
$\delta>0$ and take $n$ large. Then there is a disk
$D_0=D(x_0,n^{-2\delta})$ with the property that $\mu(D_0)\les
n^{1-2\delta}$. Write
\[ u(z) = u_1(z)+u_2(z) = \int_{D_0} \log|z-\zeta|\, \mu(d\zeta) +
\int_{\IC\setminus D_0} \log|z-\zeta|\, \mu(d\zeta)
\]
Set $D_1=D(x_0,n^{-3\delta})$. Then
\[ |u_2(z)-u_2(z')|\les n^{1-\delta} \quad \forall z,z'\in D_1\]
Cartan's theorem applied to $u_1(z)$ yields disks
$\{D(z_j,r_j)\}_j$ with $\sum_j r_j \les \exp(-2n^\delta)$ and so
that
\[ u_1(z) \gtrsim -n^{1-\delta} \qquad \forall z\in\IC\setminus
\bigcup_j D(z_j,r_j) \] Since also $u_1\le0$ on $D_1$ as well as
$|h(z)-h(z')|\les n|z-z'|$, it follows that
\begin{equation}
\label{eq:small_dev} |u(z)-u(z')|\les n^{1-\delta} \quad \forall z,z'\in D_1\setminus
\bigcup_j D(z_j,r_j)
\end{equation}
>From the Diophantine property (with $a<2$), for any $x,x'\in\tor$
there are positive integers $k,k'\les n^{4\delta}$ such that
\[ x+k\omega, x'+k'\omega \in D_1 \quad \mod \ZZ\]
In order to avoid the Cartan disks $\bigcup_j D(z_j,r_j)$ we need
to remove a set $\cB\subset \tor$ of measure $\les
\exp(-n^{\delta})$. Then from the almost
invariance~\eqref{eq:ainv}, for any $x,x'\in\tor\setminus\cB$,
\[ |u(x)-u(x')| \les n^{4\delta}+n^{1-\delta} \les n^{1-\delta}\]
This implies \eqref{eq:LDT1} with $\sigma=\tau=\delta$ and we are
done.
\end{proof}

This proof generalizes to other types of dynamics as well. For
example, let $Tx=x+\omega\;\;\mod\ZZ^d$, $d\ge2$, be a higher-dimensional shift.

\begin{defi}\label{cardef} Let $0<H<1$. For any subset $\bad\subset\Compl$ we say that $\bad\in\Car_1(H)$ if $\bad\subset\bigcup_j D(z_j,r_j)$ with
\be\label{sumradii} \sum_j r_j\le C_0\,H.\ee
If $d$ is a positive integer greater than one and $\bad\subset\Compl^d$ we define inductively that $\bad\in\Car_d(H)$ if there exists some $\bad_0\in\Car_{d-1}(H)$ so that
\[ \bad=\{(z_1,z_2,\ldots,z_d):(z_2,\ldots,z_d)\in\bad_0\mbox{\ \ or\ }z_1\in\bad(z_2,\ldots,z_d)\mbox{\ \ for some\ }\bad(z_2,\ldots,z_d)\in\Car_1(H)\}.\]
We refer to the sets in $\Car_d(H)$ for any $d$ and $H$ collectively as Cartan sets.
\end{defi}

Using the following theorem from \cite{GS1} the previous proof of~\eqref{eq:LDT1} easily generalizes.
We state the case $d=2$, with $d>2$ being similar (see also~\cite{Sch}).

\begin{theorem}\label{Cartan2} Let $u$ be a continuous function on $D(0,2)\times D(0,2)\subset\Compl^2$ so that $|u|\le 1$. Suppose further that
\[\left\{  \begin{array}{ccl} z_1\mapsto u(z_1,z_2) & \mbox{is subharmonic for each} & z_2\in D(0,2)\\
                              z_2\mapsto u(z_1,z_2) & \mbox{is subharmonic for each} & z_1\in D(0,2).
           \end{array}
  \right.
\]
Fix some $\gamma\in(0,\half)$. Given $r\in(0,1)$ there exists a polydisk $\Pi=D(\xnull_1,r^{1-\gamma})\times D(\xnull_2,r)\subset D(0,1)\times D(0,1)$ with $\xnull_1,\xnull_2\in[-1,1]$ and a set $\bad\in \Car_2(H)$ so that
\be && |u(z_1,z_2)-u(z_1',z_2')|<C_\gamma\,r^{1-2\gamma}\log\frac{1}{r}\mbox{\ \ for all\ \ }(z_1,z_2),(z_1',z_2')\in\Pi\setminus\bad\label{udevreal}\\
    && H=\exp\Bigl(-r^{-\gamma}\Bigr).\label{Heta}
\ee
\end{theorem}

The point of this theorem is that it takes the place of \eqref{eq:small_dev} in the
previous proof.

\subsection{Fourier series}

An alternative approach to \eqref{eq:LDT1} is based on Fourier
series. Indeed, one writes
\[ u(x) -\la u\ra = \frac{1}{k} \sum_{j=1}^k u(x+j\omega)-\la u\ra + O(k) =
 \sum_{\nu\ne0} \hat{u}(\nu) e(x\nu) \frac{1}{k}\sum_{j=1}^k e(j\nu\omega) + O(k)
\]
Then one has that
\[ \Bigl| \frac{1}{k}\sum_{j=1}^k e(j\nu\omega) \Bigr| \les
\min(1,k^{-1}\|\nu\omega\|^{-1})
\]
for all $\nu\ge1$. Also, it follows from~\eqref{eq:riesz} that
$|\hat{u}(\nu)|\les n|\nu|^{-1}$ which in turn implies that
\[
 |u(x)-\la u\ra| \les \frac{1}{k}\sum_{j=1}^k \Bigl|\sum_{|\nu|> K}
 \hat{u}(\nu)e(\nu (x+k\omega))\Bigr| +
\sum_{0<|\nu|\le K} n|\nu|^{-1}
\min(1,k^{-1}\|\nu\omega\|^{-1})
\]
Clearly,
\[ \Bigl\| \frac{1}{k}\sum_{j=1}^k \Bigl|\sum_{|\nu|> K}
 \hat{u}(\nu)e(\nu (x+k\omega))\Bigr|\;\Bigr\|_{L^2_x} \les n\,K^{-1/2}
\]
Taking $K=e^{n^\tau}$ it follows from the Diophantine condition that
\[ \sum_{0<|\nu|\le K} n|\nu|^{-1} \min(1,k^{-1}\|\nu\omega\|^{-1})
\les nk^{-\frac12} \log K \les n^{1+\tau} k^{-\frac12}
\]
Choosing $\tau>0$ small and $k=n^{\frac12}$, say,
yields~\eqref{eq:LDT1}.

For applications related to the study of fine properties of the IDS
it turns out to be important to obtain sharp versions of~\eqref{eq:LDT1}.
The commutative model example suggests that the optimal relation is $0\le \sigma=\tau<1$.
This is indeed the case, see Section~\ref{sec:IDS} below.

\medskip This proof also generalized to higher-dimensional tori, see~\cite{BG}
as well as~\cite{Bou2}. S.~Klein~\cite{Kle} has removed the
analyticity assumption and obtained estimates as in~\eqref{eq:LDT1}
for the Gevrey classes by means of Fourier methods (using
higher-order F\'ej\`er kernels) and suitable truncations of the
Fourier series.

\subsection{Other dynamics}

The arguments which we have just presented do not depend on positive
Lyapunov exponents. The situation is very different for the
skew-shift defined by $T(x,y)=(x+y,y+\omega)$ on $\tor^2$. The
point here is that $T^n(x,y)=(x+ny+n(n-1)\omega/2,y+n\omega)$
modulo $\ZZ^2$. Due to the presence of $ny$ in the first
coordinate we are faced with the problem that the Riesz mass
of the subharmonic extensions of $\log\|M_n(x,y,E)\|$ is now of
size $n^2$, at least if we consider the extensions to a fixed
neighborhood\footnote{Shrinking to a neighborhood to size
$O(n^{-1})$ reduces the Riesz-mass to $\les n$, but then there is a price to pay for the small diameter
of the neighborhood.} of $\tor^2\subset \IC^2$. Indeed, in this case $\|M_n(x,y+i\eps,E)\|$ behaves like
a product $\prod_{j=n}^1 e^{j\eps}$ which is of size $e^{C\eps n^2}$.
Thus, in this context neither of the two methods discussed so far
lead to a bound of the form~\eqref{eq:LDT1} for the skew-shift (the problem is that
these methods only gain a factor of $n^{-\delta}$ over the Riesz mass as far as the deviations
are concerned -- here we would therefore get $n^{2-\delta}$ for the deviations which is useless).

In \cite{BGS} a LDT is proved for the skew-shift but for large disorders.
This refers to the fact that the potential has to be of the form $\lambda V$
for large~$\lambda$.
The method in~\cite{BGS} proceeds by induction over the scale $n$, and the first stage
requires large~$\lambda$. The inductive step is realized by means of the avalanche
principle (see the following section)
which is a purely deterministic statement about products of $2\times2$ matrices.
Moreover, the analytic difficulty of having $n^2$ Riesz masses is circumvented by the
following splitting lemma from~\cite{BGS} (see also~\cite{Bou2}):

\begin{lemma}\label{lem:split} Suppose $u$ is subharmonic on $\ann_\rho$ (a $\rho$-neighborhood of $\tor$),
with $\sup_{\ann_\rho}|u|\le N$. Furthermore, assume that $u=u_0+u_1$, where
\be\label{split} \|u_0-\langle u_0\rangle\|_{L^\infty(\tor)}\le \eps_0 &\text{\ \ and\ \ }&\|u_1\|_{L^1(\tor)}\le \eps_1.\ee
Then for some constant $C_\rho$ depending only on $\rho$,
\be\label{bmo} \|u\|_{\bmo(\tor)}\ &\le& C_\rho \Bigl(\eps_0+\sqrt{N\eps_1}\Bigr).\ee
\end{lemma}

To apply this lemma one uses the avalanche principle to generate the splitting into $u_0$ and $u_1$
with an exponentially small $\eps_1\sim e^{-n}$. This allows for Riesz masses $N$ which are polynomially large, say $N=n^C$ as is
the case for the skew-shift.

\medskip
Finally, and in a very different vein, we would like to mention that LDTs have
also been established for the doubling dynamics $x\mapsto 2x\;\;\mod 1$ and for very small
disorder $\lambda>0$ in~\cite{BS}.
The latter is needed in order to apply the Figotin-Pastur formula, see~\cite{FP}.

Generally speaking, it remains an open problem to prove LDTs for Schr\"odinger cocylces
with potentials of the type $\lambda V(T^n x)$ for general classes of dynamics $T$, disorder $\lambda$,
as well as wider classes of potentials $V$.

\section{Positive Lyapunov exponents}

As we have already mentioned of the central problem concerning \eqref{eq:1} is to decided whether or not $L(E)>0$.
In the case of random i.i.d.~potentials this was established by F\"urstenberg~\cite{Fur}.
In case of quasi-periodic potentials, the well-known Herman's method~\cite{herman}
establishes this positivity for large disorders
provided the potential function $V$ is a trigonometric polynomial. Sorets and Spencer extended
this to analytic $V$. Here we present a different approach, which is based on the
following avalanche principle (AP) from~\cite{GS1} (for this version which does not
assume that the matrices belong to $SL(2,\IR)$ see \cite{GS2}).

\begin{prop}
\label{prop:AP}
Let $A_1,\ldots,A_n$ be a sequence of  $2\times 2$--matrices whose determinants satisfy
\begin{equation}
\label{eq:detsmall}
\max\limits_{1\le j\le n}|\det A_j|\le 1.
\end{equation}
Suppose that
\be
&&\min_{1\le j\le n}\|A_j\|\ge\mu>n\mbox{\ \ \ and}\label{large}\\
   &&\max_{1\le j<n}[\log\|A_{j+1}\|+\log\|A_j\|-\log\|A_{j+1}A_{j}\|]<\frac12\log\mu\label{diff}.
\ee
Then
\begin{equation}
\Bigl|\log\|A_n\cdot\ldots\cdot A_1\|+\sum_{j=2}^{n-1} \log\|A_j\|-\sum_{j=1}^{n-1}\log\|A_{j+1}A_{j}\|\Bigr|
< C\frac{n}{\mu}
\label{eq:AP}
\end{equation}
with some absolute constant $C$.
\end{prop}

The meaning of \eqref{diff} is that adjacent matrices do not cancel pairwise, whereas~\eqref{large}
insures that each matrix is sufficiently large. The conclusion is that the entire product has to be large
with the very precise difference bound from~\eqref{eq:AP}.

As an application of this principle, let us study the rate of convergence of $L_N(E)$ to $L(E)$ for the operator~\eqref{eq:1}.
We will assume that $L(E)>\gamma>0$ for some $E\in \IR$. Furthermore, we shall assume that there is a LDT of
the form~\eqref{eq:LDT1}. We shall make no other assumptions on the dynamics $T$.
Given a large integer $n$, define $k=C_0(\log N)^{\frac{1}{\tau}}$ (here $\tau$ is as in~\eqref{eq:LDT1}). Then
\[ M_N(x,E) = M_{k''}(T^{k'+\ell k}x,E)M_{k'}(T^{\ell k}x,E)\prod_{j=\ell}^1 M_k(T^{(j-1)k}x,E) \]
where $k/2\le k',k''<k$. In view of \eqref{eq:LDT1} there exists a set $\bad\subset X$ of measure $\les N^{-10}$, say,
so that for all $x\in X\setminus \bad$ we can apply the AP to this product. This requires making $C_0=C_0(\gamma)$ large.
In order to check~\eqref{diff} one uses the fact that $L_N(E)\to L(E)$ as $n\to\infty$. We can now average~\eqref{eq:AP}
over $x\in X$ which yields
\begin{equation}
\label{eq:2diff}
|L_N(E) - 2L_{2k}(E) + L_k(E)| \les \frac{(\log N)^{\frac{1}{\tau}}}{N}
\end{equation}
Applying the same reasoning with $2N$ and the same choice of $k$ yields
\[ |L_{2N}(E) - 2L_{2k}(E) + L_k(E)| \les \frac{(\log N)^{\frac{1}{\tau}}}{N}   \]
and thus also
\[ |L_{2N}(E) - L_N(E)| \les \frac{(\log N)^{\frac{1}{\tau}}}{N}   \]
Passing to the limit therefore implies that
\[ 0\le L_{N}(E) - L(E)\les \frac{(\log N)^{\frac{1}{\tau}}}{N}.  \]
This can be further improved to
\[ 0\le L_{N}(E) - L(E)\les N^{-1} \]
Moreover, this convergence holds uniformly in the energy for all
intervals $I\subset\IR$ for which $\inf_{E\in I} L(E)>\gamma>0$, see~\cite{GS1}.

\medskip
The AP can also be used to establish positive Lyapunov exponents.
Indeed, let $V:\tor^d\to \IR$ be an analytic potential and $T:\tor^d\to\tor^d$ be ergodic.
The matrix
\[ M_n(x,\lambda,E)=\prod_{j=n}^1 \bm \lambda V(T^j x)-E & -1 \\
1 & 0 \endm
\]
denotes the transfer  matrix of the equation~\eqref{eq:1} where the potential is now written as $\lambda V(T^n x)$. As before,
\[ L_n(\lambda,E)=\frac{1}{n}\int_{\tor^d}\log\|M_n(x,\lambda,E)\|\,dx \]
and $L(\lambda,E)=\lim_{n\to\infty} L_n(\lambda,E)$ exists. Finally, let $S(\lambda,E)$ be a number satisfying
\be\label{SlambdaE} S(\lambda, E) &\asymp& \sup_{n\geq1}\sup_{x\in\tor^d}\frac{1}{n}\log\|M_n(x,\lambda,E)\|.\ee
Then the following is shown in~\cite{GS1}: {\em If the weak large deviation theorem (with some $\sigma>0$)
\be\label{L1dev} \int_{\tor^d}\Bigl|\frac{1}{n}\log\|M_n(x,\lambda,E)\|-L_n(\lambda,E)\Bigr|\,dx &\le& C\,S(\lambda,E)n^{-\sigma}\ee
holds for all $n=1,2,\ldots$, then }
\[ \inf_{E}L(\lambda,E)>0\mbox{\ \ for all \ \ }\lambda>\lambda_0(V,d,\sigma).\]
This is proved inductively, the main step being described by the following lemma:

\begin{lemma} Suppose that \eqref{L1dev} holds for all $n$ with some choice of $\sigma>0$.
Then there exists a positive integer $\ell_0=\ell_0(\sigma)$ such that if
\be\label{initial} &&L_{\ell}>S(E,\lambda)\ell^{-\sigma/4}\mbox{\ \ and\ }L_{\ell}(E,\lambda)-L_{2\ell}(E,\lambda)<\frac{L_\ell(E,\lambda)}{8}\ee
for some $\ell\ge \ell_0$, then $L(E,\lambda)>L_\ell(E,\lambda)/2$.
\label{plusforcer}\end{lemma}

This lemma does not require any further information about $T$ or $V$ other than~\eqref{L1dev}.
On the other hand, to insure that the conditions of this lemma are met, one chooses $\lambda$ large
using analyticity of~$V$. It would be interesting to apply this method to non-analytic $V$ which
satisfy some natural non-degeneracy assumption (in particular, one would need to establish~\eqref{L1dev}).

\smallskip K.\ Bjerkl\"ov~\cite{Bje} recently used this result to prove positive exponents for some interval
of energies for his class of potentials which exhibit mixed behavior (i.e., both zero and positive exponents).

\section{Regularity of the IDS}
\label{sec:IDS}

Let us now assume that we have the following sharp LDTs for $u_n(x)=\log\|M_n(x,E)\|$:
\begin{equation}
\label{eq:LDT2}
 \nu\bigl(\{x\in X: |u_n(x)-nL_n(E)|>n\delta\}\bigr)\leq \exp\Bigl(-c(\delta)n+C(\log n)^A\Bigr).
\end{equation}
where $c(\delta)>0$ and $C,A$ are some constants\footnote{We can consider these estimates as a ``black box'' without
specifying $T$ or $V$ further}.
For the case of $X=\tor$ and the shift by a Diophantine\footnote{This means that $\|n\omega\|\ge \frac{c(\omega)}{n(\log n)^a}$ with $a>1$}
 $\omega$, such estimates were obtained in~\cite{GS1}
with $c(\delta)=\delta^2$. Assuming positive Lyapunov exponents $L(E)>\gamma$ one can further show that $c(\delta)=c(\gamma)\delta$
which shows that one can take $0\le\sigma=\tau<1$ in~\eqref{eq:LDT1} (the latter is done via the AP).

Then by the arguments of the previous section we obtain the following stronger version of~\eqref{eq:2diff}
\[ |L_N(E) - 2L_{2k}(E) + L_k(E)| \les \frac{\log N}{N}  \]
This is due to the fact that we can break up $M_N$ into products of matrices of size $k=[C\log N]$ when applying the AP.
Therefore,
\begin{align*}
 |L(E)-L(E')| &\le |L_N(E)-L(E)| + |L_N(E')-L(E')|  \\
& \quad +|L_N(E) - 2L_{2k}(E) + L_k(E)| + |L_N(E') - 2L_{2k}(E') + L_k(E')| \\
& \quad +2|L_{2k}(E)-L_{2k}(E')| + |L_k(E)-L_k(E')| \\
&\les   N^{-1}\log N\, + |E-E'| \exp(Ck) \les   N^{-1}\log N\, + N^B |E-E'|
\end{align*}
Consequently, choosing $N=|E-E'|^{-\alpha}$ for some $0<\alpha<1$, we deduce that $L(E)$ and therefore also that the IDS
(by~\eqref{eq:thou}) are H\"older continuous.

This approach to the regularity of $L(E)$ was subsequently modified by other authors. For example, Bourgain
and Jitomirskaya~\cite{BouJit} use this very approach from~\cite{GS1} to show that $L(E,\omega)$ is jointly continuous
away from rational~$\omega$ for the case of the shift on~$\tor$. Their argument is based on sharp large deviation theorems,
the avalanche principle, as well as the difference relation~\eqref{eq:2diff}.

\section{Eigenvalues, localization, and the zeros of $f_N(z,\omega,E)$ in $z$}

In \cite{GS2} a different approach to the regularity of the IDS was developed which allows us to obtain a lower
bound on the H\"older exponent by non-perturbative methods. In fact, for the almost Mathieu model with $\lambda>2$
it is shown that the H\"older regularity is~$\frac12-\eps$ for any $\eps>0$.  Moreover, off a set of Hausdorff dimension
zero the IDS is $1-\eps$ H\"older regular for any $\eps>0$ (the latter requires the removal of a set of $\omega$ of
measure zero). Similar results are obtained for other potentials
assuming positive Lyapunov exponents as well as the strong Diophantine condition~\eqref{eq:strong_dio}.

It is well-known that in the case of the almost Mathieu model the exponent
$\frac12$ cannot be improved. For large disorders this was observed by Sinai~\cite{Si1},
whereas Puig~\cite{Pui1}, \cite{Pui2} has obtained this non-perturbatively. On the other hand, Bourgain~\cite{Bou1}
has shown for the almost Mathieu operator that the H\"older
exponent is no worse than $\half-\eps$ for very large disorders.

In the almost Mathieu case the optimality of $\frac12$ is intimately connected with the Cantor structure of
the spectrum. In fact, it is at the gap edges that one encounters loss of the Lipschitz behavior.
For more general potentials as in~\cite{GS2} this connection is not clear and it would certainly be of great interest
to elucidate the connection between gaps and the regularity of the IDS further.

We now set out to describe some of the basic ingredients of~\cite{GS2}.

\subsection{Large deviation theorems for the entries}

Recall that
\[
M_n(x,E) = \left[\begin{array}{cc} f_n(x,E) & -f_{n-1}(Tx,E) \\
                                 f_{n-1}(x,E) & -f_{n-2}(Tx,E) \\
                                     \end{array} \right]
\]
where
\begin{equation} \label{eq:fn_def}
f_n(x,E) = \det\left[
\begin{array}{ccccccccc}
                         v(1,x)-E & -1 & 0 & 0 & . & . & . & . &  0    \\
                        -1 &  v(2,x)-E & -1 & 0 & 0 & . & . & . &  0 \\
                        0 & -1 &  v(3,x)-E & -1 & 0 & 0 &  . & . & 0 \\
                        . & . & . & . & . & . & . & . & . \\
                        . & . & . & . & . & . & . & . & . \\
                        . & . & . & . & . & . & . & . & . \\
                        . & . & . & . & . & . & . & . & . \\
                        0 & 0 & . & . & . & . & . & -1 &  v(n,x)-E
\end{array} \right].
\end{equation}
It is customary to denote the matrix on the right-hand side as $H_{[1,n]}(x)-E$ so that one as
$f_n(x,E)=\det(H_{[1,n]}(x)-E)$. The following result is proved in~\cite{GS2}. Henceforth, we shall
assume that the Lyapunov exponents are positive as well as that $\omega$ is strongly Diophantine.

\begin{lemma}
\label{lem:det_bmo}
There exist constants $A$ and $C$
depending on $\omega$ and the potential $V$, so that for every $n\ge1$
\be
\label{eq3.35}
&& \Big| \int_0^1 \log |\det (H_{[1,n]} (x,\omega) - E) |\, dx
- n \, L_n (\omega,E) \Big| \le C \\
\label{eq3.36}
&& \|  \log | \det (H_{[1,n]} (x,\omega) - E) |\, \|_\BMO \le
C(\log n)^A.
\ee
In particular, for every $n\ge1$,
\begin{equation}\label{eq:det_LDT}
\mes \Big[ x\in \TT \:\big|\:
|\log | \det (H_{[1,n]} (x,\omega) - E) | - n\, L_n (\omega,E) |
> H\Big] \le C\exp \left( - \frac{cH}{(\log n )^A} \right)
\end{equation}
for any $H>(\log n)^A$. Moreover, the set on the left-hand side is contained in at most $\les n$ intervals
each of which does not exceed the bound stated in~\eqref{eq:det_LDT} in length.
\end{lemma}

The point here is of course that the entries of $M_n$ satisfy the same bound as $M_n$ itself.
One basic step in the proof of this lemma is to show that
\begin{equation}
\label{eq:mean} \int_{\tor} \log |f_n(x,E)|\, dx \ge nL_n(E) - Cn^\sigma
\end{equation}
with some $\sigma<1$. Then combine this with a uniform upper bound of the form (see~\cite{GS1})
\[ \sup_{x\in\tor} \log |f_n(x,E)|\le \sup_{x\in\tor} \log \|M_n(x,E)\| \le nL_n(E) + Cn^\sigma \]
to conclude that
\[ \log |f_n(x,E)| = \int \log |f_n(x,E)| \,dx  + u_1(x) \text{\ \ with\ \ }\|u_1\|_1 \les n^{\sigma} \]
Thus, by Lemma~\ref{lem:split} we conclude that
\[ \big\| \log |f_n(x,E)| \big \|_{\bmo} \les n^{(1+\sigma)/2} \]
which beats the trivial bound of $n$. We used here that the Riesz mass of the subharmonic extension of
$\log |f_n(x,E)|$ to a neighborhood of the circle is $\les n$.

The John-Nirenberg inequality therefore implies the large deviation theorem
\begin{equation}
\label{eq:weak_LDT} \mes(\{ x\in\tor\::\: |\log|f_n(x,E)|- nL_n(E) | > n^{1-\delta}\}) \les \exp(-n^\delta)
\end{equation}
provided $\delta>0$ is sufficiently small.

This is of course considerably weaker than Lemma~\ref{lem:det_bmo}. To improve on it, we
apply the avalanche principle to the product
\[
\left[ \begin{array}{cc} 1&0\\ 0&0 \end{array} \right]
M_{n} (x,\omega,E)
\left[ \begin{array}{cc} 1&0\\ 0&0 \end{array} \right]
= \left[ \begin{array}{cc} f_{n} (x,E) &0\\ 0&0 \end{array}\right].
\]
Note that the AP does not require $SL(2,\IR)$ matrices, but rather~\eqref{eq:detsmall}
which holds here. Write  $n=\ell_1+(m-2)\ell+\ell_m$ where $\ell \asymp (\log n )^{C_0} $,
$\ell_1  \asymp \ell_n \asymp \ell $, and set $s_1 = 0$, $s_j = \ell_1 + (j-2) \ell $ for $2\le j \le m$. Hence,
\[ \left[ \begin{array}{cc} f_{n} (x,E) &0\\ 0&0 \end{array}\right] = \prod_{j=m}^1 A_j(x) \]
where $A_j (x) = M_\ell (x+s_j \omega) $, for $2\le j \le m-1$, and $A_1(x)= M_{\ell_1} (x) \bm 1&0\\0&0 \endm$,
$A_m(x)= \bm 1&0\\0&0 \endm M_{\ell_m} (x+s_m\omega)$.
One checks easily from \eqref{eq:weak_LDT} that the  conditions~\eqref{large} and~\eqref{diff}
hold up to a set of $x$ of measure $< n^{-100}$, say. Hence, by Proposition~\ref{prop:AP},
\begin{equation}
\label{eq3.40}
\log |f_n (x,E) | = - \sum\limits_{j=2}^{m-1} \log \| A_j (x) \|
+ \sum\limits_{j=1}^{m-1} \log\| (A_{j+1} A_j) (x) \| + O\left( \frac1{n}\right).
\end{equation}
We now invoke the following LDT for sums of shifts of subharmonic functions,
see~Theorem~3.8 in \cite{GS1}: For any subharmonic function $u$ on a neighborhood of $\tor$
 with bounded Riesz mass and harmonic part
\begin{equation}
\label{eq:sumsofshifts}
 \mes\Bigl[x\in\tor \:|\: \bigl|\sum_{k=1}^n u(x-k\omega)- n\langle u\rangle\bigr|>\delta n\Bigr]<\exp(-c\delta n+r_n)
\end{equation}
where $r_n \lesssim (\log n)^A$.
The sums in~\eqref{eq3.40} involve shifts by $\ell\omega$  rather than~$\omega$. In order to overcome this,
note that we can take $\ell_n > 2\ell$, say. Repeating the argument that lead to~\eqref{eq3.40} $\ell-1$ times
with the length of $A_1$ increasing by one and that of~$A_m$ decreasing by one, respectively,  at each step leads to
\be
\nn
 \log | f_n (x,\omega,E) |
&=& - \frac{1}{\ell}\sum_{k=0}^{\ell-1}\sum\limits_{j=2}^{m-1}
\log \| A_j (x+k\omega) \| + \frac{1}{\ell} \sum_{k=0}^{\ell-1} \sum\limits_{j=2}^{m-2} \log\| (A_{j+1} A_j) (x+k\omega) \|
+ \frac{1}{\ell}\sum_{k=0}^{\ell-1} u_k(x) + O\left( \frac1{n}\right) \nn \\
&=& - \frac{1}{\ell}\sum\limits_{j=\ell}^{(m-1)\ell-1} \log \| M_\ell (x+j\omega) \| +
 \frac{1}{\ell}\sum\limits_{j=\ell}^{(m-1)\ell-1} \log\| M_{2\ell} (x+j\omega) \|
+ \frac{1}{\ell}\sum_{k=0}^{\ell-1} u_k(x) +     O\left( \frac1{n}\right). \label{eq:shift_rep}
\ee
The functions $u_k$ compensate for omitting the terms $j=1$ and $j=m-1$ when
summing~$\log \| A_{j+1} A_j \|$.  They are subharmonic, with Riesz mass and harmonic part bounded by~$(\log n )^{C_0}$.
Estimating the sums involving~$M_\ell$ and~$M_{2\ell}$ by means of~\eqref{eq:sumsofshifts},
and the sums involving~$u_k$ directly by means of Cartan's bound shows that
there exists $\cB\subset\tor$ of measure $\le \exp (-(\log N)^{C_0})$, so that for all $x\in\tor\setminus\cB$,
\begin{equation} \nn
\Big| \log | f_n (x,\omega,E) | -
\langle \log |f_n (x,\omega,E)| \rangle \Big|
\le (\log n )^{2C_0}.
\end{equation}
Thus,
\begin{equation}\nn
\log | f_n  (x,\omega,E) | = u_0 (x) + u_1 (x) \ ,
\end{equation}
where
\[ \| u_0 - \langle \log | f_n(\cdot , E) | \rangle \|_{L^\infty(\tor)}  \le (\log n)^{2C_0},\]
and
\begin{eqnarray}\nn
\| u_1 -\langle \log | f_n (\cdot , E) |
\rangle \|_{L^1 (\tor) }
&\lesssim & \| \log | f_n(\cdot , E) |\;  \|_{L^2 (\tor)} \sqrt{\mes (\cB)} \\
&\lesssim &  n \cdot \sqrt{\mes (\cB)}
\lesssim  \exp \left( - \frac14 (\log n)^{C_0} \right).\nn
\end{eqnarray}
Applying Lemma~\ref{lem:split}
 one now obtains that
\begin{eqnarray}\nn
\Big\| \log | f_n (x,\omega,E) | \Big\|_{\BMO (\tor)}
&\le& C \left( (\log n)^{2C_0 + 1} +
\sqrt{n \cdot \exp \left( -\frac14 (\log n)^{C_0}\right)} \right) \\
&\le& C\,(\log n)^{2C_0 + 1} \ , \nonumber
\end{eqnarray}
as claimed.

It therefore remains to obtain~\eqref{eq:mean}. For this, as well as
other details of Lemma~\ref{lem:det_bmo} we refer the reader to Section~2 of~\cite{GS2}.

\medskip We remark that \eqref{eq:shift_rep} illustrates how the AP allows us to write
the determinants $f_n(x,E)$ as rational functions which are composed of products of shifts
of ``short'' functions (more precisely, of functions with small Riesz mass).
This cannot be done for all $x$ (because of the bad sets in the LDTs)
and also leads to certain small errors. This approximate factorization is one of the basic
tools of~\cite{GS2}.

\subsection{Uniform upper bounds and zeros of determinants}

The following result based on Lemma~\ref{lem:det_bmo} improves on
these uniform upper bounds. Uniform upper bounds on the norm of the
monodromy matrices in terms of $L(E)$ were found in \cite{BG},
\cite{GS1}. The $(\log N)^A$ error obtained in~\cite{GS2} (rather
than $N^\sigma$, say, as in~\cite{BG} and~\cite{GS1}) is crucial for
the study of the fine properties of the integrated density of
states.

\begin{lemma}\label{lem:5.2}
Let $\w$ be as in~\eqref{eq:strong_dio}. Assume $L(\w, E) > 0$.
Then for all large integers $N$,
$$
\sup_{x\in \tor}\, \log \|M_N\xwe\| \le NL_N\we + C(\log N)^A\ ,
$$
for some constants $C$ and $A$.
\end{lemma}
\begin{proof} We only consider $x$ and suppress $\w$ and $E$ from most of the notation.
Take $\ell\asymp (\log N)^A$.  Write $N = (n -1)\ell +r$, $\ell \le r < 2\ell$ and correspondingly
$$
M_N(x) = M_r\bigl(x + (n-1)\ell\w\bigr) \prod^0_{j=n-2}\, M_\ell(x + j\ell \w)\ .
$$
The avalanche principle and the LDT~\eqref{eq:LDT2} imply that for
every small $y$ there exists $\cB_y \subset \tor$ so that
$\mes(\cB_y) < N^{-100}$ and such that for $x \in [0, 1]\setminus
\cB_y$, \beeq
\begin{aligned}
\label{eq:5.21}
\log \|M_N(x + iy)\| & = \sum^{n-3}_{j=0}\, \log \|M_{2\ell}(x + j\ell\w + iy)\| - \sum^{n-2}_{j=1}\, \log \|M_\ell(x + j\ell\w + iy)\|\\
&\qquad + \log \|M_r\bigl(x + (n-1)\ell\w\bigr) M_\ell\bigl(x + (n-2)\ell\w\bigr)\| + O(1)\\
& = \sum^{n-3}_{j=0}\, \log \|M_{2\ell}(x + j\ell\w + iy)\| - \sum^{n-2}_{j=1}\, \log \|M_\ell(x + j\ell\w + iy)\| + O(\ell)\ .
\end{aligned}
\eneq
Combining the elementary almost invariance property
$$
\log \|M_N(x + iy)\| = \ell^{-1} \sum_{0 \le j \le \ell-1}\, \log \|M_N(x + j\w + iy)\| + O(\ell)
$$
with (\ref{eq:5.21}) yields
\beeq\label{eq:5.22}
\begin{aligned}
\log \|M_N(x+iy)\| & = \ell^{-1} \sum_{0 \le j < N}\, \log \|M_{2\ell}(x + j\w + iy)\|\\
&\qquad -\ell^{-1} \sum_{0 \le j < N}\, \log \|M_\ell(x + j\w + iy)\| + O(\ell)\ ,
\end{aligned}
\eneq for any $x \in [0, 1]\setminus\cB'_y$, where $\mes\cB'_y <
N^{-9}$.  Integrating (\ref{eq:5.22}) over $x$ shows that
\beeq\label{eq:5.23} L_N(y, E) = 2L_{2\ell}(y, E) - L_\ell(y, E) +
O(\ell/N)\ . \eneq This identity is formula (5.3) in \cite{GS1}
(with $y = 0$). Since the Lyapunov exponents are Lipschitz in $y$,
the sub-mean value property of subharmonic functions on the disk
$\cD(x, 0; \delta)$ with $\delta = N^{-1}$ in conjunction with
(\ref{eq:5.22}) and (\ref{eq:5.23}) implies that, for every $x \in
\tor$,
\beeq\label{eq:5.24}
\begin{aligned}
& \log \|M_N(x)\| - \int^1_0 \log \|M_N(\xi)\|d\xi\\
&\le \strichint_{\cD(x,0;\delta)} \Biggl[\sum_{0 \le j < N}\, u(\xi + j\w + i\eta) - N\langle u(\cdot + i\eta)\rangle\Biggr]d\xi\, d\eta\\
& - \strichint_{\cD(x,0;\delta)}\Biggl[\sum_{0 \le j < N}\, v(\xi + j\w + i\eta) - N\langle v(\cdot + i\eta)\rangle\Biggr]d\xi\, d\eta + O(\ell)\ ,
\end{aligned}
\eneq where $\strichint_{\cD(x,0;\delta)}$ denotes the average over
the disk, \beeq\nn u(\xi + i\eta) := \ell^{-1} \log \|M_{2\ell}(\xi
+ i\eta)\|\quad \text{and}\quad v(\xi + i\eta) := \ell^{-1} \log
\|M_\ell(\xi + i\eta)\|\ , \eneq and $\langle\cdot\rangle$ denotes
averages over the real line.  The lemma now follows easily
from~\eqref{eq:5.24}.
\end{proof}

The first application of this estimate is as follows:

\begin{lemma} Let $\w$ satisfy (\ref{eq:strong_dio}).  Then for any $x_0 \in \tor$, $E_0 \in \IR$ one has
\begin{gather}
\label{eq:5.26} \#\left\{E \in \IR: f_N\bigl(e(x_0), \w, E\bigr) = 0,\ |E - E_0| < \exp\bigl(-(\log N)^A\bigr)\right\} \le (\log N)^{A_1}\\[6pt]
\label{eq:5.27} \# \left\{z \in \IC: f_N(z, \w, E_0) = 0,\ |z -
e(x_0)| < N^{-1}\right\} \le (\log N)^{A_1}
\end{gather}
for all sufficiently large $N$.
\end{lemma}

\begin{proof} It follows from Lemma~\ref{lem:det_bmo} that
\begin{equation*}
\begin{aligned}
\sup \Bigl\{\log \big |f_N\bigl(e(x),\w, E\bigr)\big| : x \in \tor,\ E \in \IC,\ |E - E_1| & < \exp\bigl(-(\log N)^A\bigr)\Bigr\}\\
& \le NL_N(\w, E_1) + (\log N)^B
\end{aligned}
\end{equation*}
for any $E_1$.  Due to the large deviation theorem, there exist $x_1, E_1$ such that $|x_0 - x_1| < \exp \bigl(-(\log N)^{2A}\bigr)$, $|E_0 - E_1| < \exp\bigl(-(\log N)^{2A}\bigr)$ so that
$$
\log \big |f_N\bigl(e(x_1),\w, E_1\bigr)\big | > NL_N(\w, E_1) - (\log N)^{4A}\ .
$$
Due to Jensen's well-known formula, see \eqref{eq:jensen} below,
$$
\#\left\{E: f_N \bigl(e(x_1),\w, E\bigr) = 0,\ |E - E_1| < \exp \bigl(-(\log N)^A\bigr) \right\} \le (\log N)^C\ .
$$
Since $\big \|H_N^{(D)}(x_0, \w) - H_N^{(D)}(x_1,\w) \big \| \lesssim \exp \bigl(-(\log N)^{2A}\bigr)$ and since $H_N^{(D)}(x_0, \w)$ is self adjoint one has
\begin{align*}
& \#\left(E: f_N\bigl(e(x_0),\w, E\bigr)  = 0,\ |E - E_0| < \exp\bigl(-\log N)^{2A}\bigr)\right\}\\
& \le \#\left\{E: f_N\bigl(e(x_1),\w, E\bigr) = 0,\ |E - E_1| < \exp \bigl(-(\log N)^A\bigr) \right\} \le (\log N)^C\ .
\end{align*}
That proves (\ref{eq:5.26}).  The proof of (\ref{eq:5.27}) similar.
\end{proof}

These estimates of the local number of zeros of the determinants
$f_N$ allows one to factorize $f_N$ in each neighborhood of size
$\exp\bigl(-(\log N)^{A_1}\bigr)$ using the Weierstrass preparation
theorem, with a polynomial factor of degree at most $(\log N)^C$.
For example, the following is proved in~\cite{GS2}, see Section~6.

\begin{prop}
\label{prop:fN_prep_z}
 Given $z_0 \in \cA_{\rho_0/2}$, $E_0 \in \IC$, and $\omega_0$ as in~\eqref{eq:strong_dio}, there
exist a polynomial
\[  P_N(z, \omega,E) = z^k + a_{k-1} (\omega,E) z^{k-1} + \cdots + a_0(E,
\omega)\]
 with $a_j(\omega,E)$ analytic in $\cD(E_0, r_1)\times \cD(\omega_0, r_1)$, $r_1 \asymp \exp
\left(-(\log N)^{A_1}\right)$ and an analytic function \[g_N(z,
\omega,E),\quad (z, \omega,E) \in \cP = \cD(z_0, r_0) \times
\cD(E_0, r_1) \times \cD(\omega_0,r_1)\] with $r_0 \asymp N^{-1}$
such that:
\begin{enumerate}
\item[(a)] $f_N(z, \omega,E) = P_N(z, \omega,E) g_N(z, \omega,E)$

\item[(b)] $g_N(z, \omega,E) \ne 0$ for any $(z, \omega,E) \in \cP$

\item[(c)] For any $(\omega,E) \in  \cD(\omega_0, r_1)\times \cD(E_0, r_1) $, the
polynomial
 $P_N(\cdot, \omega,E)$ has no zeros in $\IC \setminus \cD(z_0, r_0)$

\item[(d)] $k = \degg P_N(\cdot, \omega,E) \le (\log N)^A$.
\end{enumerate}
\end{prop}

Another application of the uniform upper estimates is the following
analogue of Wegner's estimate from the random case (see \cite{We}).
It will be important that there is only a loss of $(\log N)^A$
in~(\ref{eq:5.28}).

\begin{lemma} Suppose $\w$ satisfies \eqref{eq:strong_dio}.
Then for any $N \gg 1$, $E \in \IR$, $H \ge (\log N)^A$ one has
\beeq \label{eq:5.28} \mes \left\{x \in \tor: \dist \bigl(\rsp
H_N\xw, E\bigr) < \exp(-H)\right\} \le \exp \bigl(-H/(\log
N)^A\bigr)\ . \eneq Moreover, the set on the left-hand side is
contained in the union of $\lesssim N$ intervals each of which does
not exceed the bound stated in (\ref{eq:5.28}) in length.
\end{lemma}

\begin{proof} By Cramer's rule
$$
\Bigl|\bigl(H_N\xw - E\bigr)^{-1}(k, m) \Bigr|= {\big|f_{[1, k]}\bigl(e(x), \w, E\bigr)\big |\,\big |f_{[m+1, N]}\bigl(e(x), \w, E\bigr)\big |\over \big |f_N\bigl(e(x), \w, E\bigr)\big |}\ .
$$
By Lemma 5.2
$$
\log \big |f_{[1, k]}\bigl(e(x), \w, E\bigr)\big | + \log \big |f_{[m+1, N]}\bigl(e(x), \w, E\bigr)\big | \le NL\we + (\log N)^{A_1}
$$
for any $x \in \tor$.  Therefore,
$$
\big\| \bigl(H_N\xw - E\bigr)^{-1}\big \| \le N^2\, {\exp\bigl(NL\we + (\log N)^A\bigr)\over \big| f_N\bigl(e(x),\w, E\bigr)\big |}
$$
for any $x \in \tor$.  Since
$$
\dist\bigl(\rsp\bigl(H_N\xw, E\bigr) = \big \|\bigl(H_N\xw -
E\bigr)^{-1}\big \|^{-1}\ ,
$$
the lemma follows.
\end{proof}


\subsection{Elimination of resonances and the separation of zeros}

Given arbitrary $E \in \IR$, the typical distance from $E$ to the
eigenvalues of equation (\ref{eq:1}) on a finite interval $[-N, N]$
should be at least~$\const\cdot N^{-1}$.  If for some $E \in \IR$
and $x \in \tor$ this distance $\rho$ is considerably
smaller\footnote{Technically speaking, this means $\exp(-N^b)$ with
$b<1$} than $N^{-1}$, then we say that $(E,x)$ are {\em in
resonance} and we refer to $\rho^{-1}$ as the {\em magnitude of the
resonance}. Clearly, the $x$-averaged distribution of the
eigenvalues of (\ref{eq:1}) on the interval $[-N, N]$ controls the
probability of resonances. For more accurate estimates the fine
properties of this distribution are very important.

Assume that for some $E \in \IR$, $x \in \tor$, $n \in \IZ$, both
$(E, x)$ and $(E, T^n x)$ are in resonance.  In this case we say
that this pair forms a {\em double resonance}.  Double resonances
play a crucial role in any proof of  Anderson localization, i.e.,
that the eigenfunctions of (\ref{eq:1}) decay exponentially as $|n|
\to \infty$. This was found in Sinai's classical work \cite{Si1},
where Anderson localization was established for (\ref{eq:1}) with
$V(x) = \cos( 2\pi x)$ and large $|\lambda|$.  A novel,
non-perturbative approach to the study of double resonances was
found by Bourgain and the first author in~\cite{BG}. It is based on
the following notion:

\begin{defi} A set $S \subset \IR^m$ is called {\em semi-algebraic} if it
is a finite union of sets defined by a finite number of polynomial inequalities.
\end{defi}

For instance, let $V(x)$, $x \in \tor$ be a trigonometric polynomial
and consider the dynamics of the shift on~$\tor$. Then for fixed $x$
the double resonances can be included into a semialgebraic set (in
the $\omega,E$ plane). An important parameter of a semialgebraic set
is its degree which equals \[\text{ (number of polynomials involved)
$\times$ (maximal degree of these polynomials)}\]  If $V$ is a
trigonometric polynomial, then the degree of the set of resonances
on the interval $[-N, N]$ is at most $N^C$ for some absolute
constant~$C$.  On the other hand, the only ``dangerous resonances''
for the localization are those of magnitude  $\exp(cN)$.  That
allows one to eliminate double resonances for the case of the shift
$x \to x + \w$ and skew-shifts $(x,y) \to (x+\w, y+x)$ using simple
geometrical ideas related to semialgebraic sets, see \cite{Bou2},
Section~9. In \cite{GS2} we develop a more quantitative method for
analyzing the resonances, which is based on the theory of resultants
and discriminants of polynomials.  The polynomials in question are
those which arise in the factorization of the determinants $f_N$ via
the Weierstrass preparation theorem in the phase variable $x$.  This
method turns the set of resonances into a set on which some analytic
function (in the case of a double resonance it is the resultant)
attains very small values. Cartan's estimate from above applied to
this analytic function then leads to bounds on the measure and
complexity of this set. The logic of this is captured by the
following lemma from Section~7 in \cite{GS2}.
 We also use the following notation: Given $\uw_0 = (w_{1,0},
\dots, w_{d,0}) \in \IC^d$, $\ur = (r_1, \dots, r_d)$, $r_i > 0$, $i
= 1, 2, \dots, d$, set
\[S_{\uw_0, \ur} (w_1, \dots, w_d) = \bigl(r_1^{-1}(w_1 - w_{1,0}),
\dots, r_d^{-1}(w_d - w_{d,0})\bigr).\]

\begin{lemma}
\label{lem:H2O} Let $P_s(z, \uw) = z^{k_s} + a_{s, k_s -1} (\uw)
z^{k_s -1} + \cdots + a_{s,0}(\uw)$, $z \in \IC$, where $a_{s,
j}(\uw)$ are analytic functions defined in some polydisk $\cP =
\prod\limits_i D(w_{i,0}, r)$, $\uw = (w_1, \dots, w_d) \in \IC^d$,
$\uw_0 = (w_{1,0}, \dots, w_{d, 0}) \in \IC^d$, $s = 1, 2$. Assume
that $k_s > 0$, $s = 1, 2$ and set $k=k_1k_2$. Suppose that for any
$\uw \in \cP$ the zeros of $P_s(\cdot, \uw)$ belong to the same disk
$D(z_0, r_0)$, $r_0 \ll 1$, $s = 1, 2$.  Let $t > 16k\, r_0\,
r^{-1}$. Given $H \gg 1$ there exists a set $\cB_H\subset \cP$ such
that $S_{\uw_0, (16kr_0t^{-1}, r,\dots, r)} (\cB_H) \in
\car_d(H^{1/d}, K)$, $K = CHk$ and for any $\uw \in \cD(w_{1,0},
8kr_0/t) \times \prod\limits^d_{j=2} \cD(w_{j,0}, r/2) \setminus
\cB_H$ one has
\begin{equation}
\label{eq:zero_dist} \dist\left(\bigl\{\mbox{zeros of $P_1(\cdot,
\uw)$}\bigr\}, \bigl\{\mbox{zeros of $P_2\left(\cdot + t(w_1 -
w_{1,0}), \uw\right)\bigr\}$}\right) \ge e^{-CHk}\ .
\end{equation}
\end{lemma}

It is instructive for the reader to first consider the meaning of
the previous lemma for the case where neither $P_1$ nor $P_2$ depend
on $\uw$. Lemma~\ref{lem:H2O} is the principal tool for eliminating
resonant phases and energies in the paper~\cite{GS2}. As a typical
application of it we mention the following lemma on the separation
of the zeros of determinants from Section~8 of~\cite{GS2}.
 Let $T(x) = x + \w$ be a shift and $f_N(z,\w, E)$ be the Dirichlet
determinants defined as in (\ref{eq:fn_def}) with $v(n, x) = V(x +
n\w)$. Also, ${\mathcal Z}(f,z_0,r_0)$ denotes the set of zeros of
$f$  in the disk ${\mathcal D}(z_0,r_0)$.

\begin{lemma}
\label{lem:5.6} Let $C_1 > 1$ be an arbitrary constant.  Given
$\ell_1 \ge \ell_2 \gg 1$, $t > \exp\bigl((\log \ell_1)^A\bigr)$, $H
\gg 1$, there exists a cover of $\tor \times \tor_{c,a} \times
[-C_1, C_1]$ by a system $\cS$ of polydisks
$$\cD(x_m, r) \times \cD(\w_m, rt^{-1})\times \cD(E_m, r),\quad x_m \in \tor,\ E_m \in [-C_1, C_1]\ ,
$$
with $\w_m \in \tor_{c,a}$, and $r = \exp\bigl(-(\log \ell_1)^{A_2}\bigr)$,
and which satisfies the following properties: $\cS$ has multiplicity $\lesssim 1$,
cardinality $\#(\cS) \lesssim t\exp \bigl((\log \ell_1)^{A_1}\bigr)$ and for each $m$,
there exists a subset $\Omega_{\ell_1, \ell_2, t, H, m}\subset \cD(\w_m, rt^{-1}/2)$ with
$$
\cS_{\w_m, rt^{-1}/2} (\Omega_{\ell_1, \ell_2, t, H, m}) \in \Car_1 (H^{1/2}, K),\quad K = (\log \ell_1)^B
$$
such that for any $\w \in \cD(\w_m, rt^{-1}/2)\setminus
\Omega_{\ell_1, \ell_2, t, H, m}$ there exists a subset
\[\cE_{\ell_1, \ell_2, t, H,\w, m} \subset \cD(E_m, r),\qquad \cS_{E_m,
r}(\cE_{\ell_1, \ell_2, t, H,\w, m}) \in \Car_1(H^{1/2}, K) \] such
that for any $E \in \cD(E_m, r)\setminus \cE_{\ell_1, \ell_2, t,
H,\w, m}$ one has
$$
\dist\Bigl(\cZ\bigl(f_{\ell_1}(\cdot, \w, E), e(x_m), r\bigr),
\cZ\bigl(f_{\ell_2}(\cdot e(t\w),\w, E), e(x_m), r\bigr)\Bigr) > e^{-H(\log \ell_1)^C}\ .
$$
\end{lemma}

\subsection{Localization}

Let $f_N(z, \w, E)$ be the Dirichlet determinants for equation
(\ref{eq:1}) on the interval $[1, N]$ with the dynamics $Tx = x +
\w$. The separation of the zeros of $f_N(\cdot, \w, E)$ and $f_N(x +
n\w, \w, E)$ described in Lemma~\ref{lem:5.6} is achieved due to the
dynamics $z \to ze(\w)$. It is shown in~\cite{GS2} that this
separation property implies that for most energies the associated
eigenfunctions on a finite interval $[-N,N]$ are exponentially
localized. Furthermore, the authors also show that this localization
on a finite interval can be used to obtain a lower bound on the
minimal distance between a typical pair of eigenvalues, i.e.,
between the zeros of $E\mapsto f_N(x, \w,E)$. This application of
Anderson localization will be described in the following subsection.
Here we outline how one can find the localized eigenfunctions.

Any solution of the equation
\begin{equation}
\label{eq:hamilton} -\psi(n+1) - \psi(n-1) + v(n)\psi(n) = E\psi(n)\
,\quad n \in \IZ\ ,
\end{equation}
obeys the relation \beeq \label{eq:poisson} \psi(m) = \cG_{[a, b]}
(E)(m, a-1)\psi(a-1) + \cG_{[a, b]} (E)(m, b+1)\psi(b+1),\quad m \in
[a, b] \eneq where $\cG_{[a,b]} (E) = \left(H_{[a,b]}
-E\right)^{-1}$ is the Green function, $H_{[a,b]}$ being the linear
operator defined by \eqref{eq:hamilton} for $n \in [a, b]$ with zero
boundary conditions.

In view of~\eqref{eq:poisson}, one can prove exponential
localization on $[1,N]$ by showing that outside of some subinterval
$[n_0-L,n_0+L]$ (where $\log L \ll \log N$) all Green functions of a
much smaller scale (say, scale $(\log N)^C$) have exponential
off-diagonal decay. Indeed, \eqref{eq:poisson} would then imply that
any eigenfunction is exponentially small outside of the window
$[n_0-L,n_0+L]$. This strategy was introduced by Fr\"ohlich and
Spencer in their fundamental work (see \cite{FS1}, \cite{FS2},
\cite{FSW}) on Anderson's model. The following simple lemma shows
that the question of exponential off-diagonal decay for the Green
function is intimately related to the LDTs for the determinants.

\begin{lemma} Let $\w \in \tor_{c,a}$.  Suppose $L(\w, E_0)=\gamma > 0$,
\beeq \label{eq:large_det} \log \big |f_\ell(z_0, \w, E_0)\big | >
\ell L(\w, E_0) - K/2 \eneq
 for some $z_0 = e(x_0)$, $x_0 \in \tor$,
$E_0 \in \IR$, $\ell \gg 1$, $K > (\log \ell)^A$.  Then
\begin{align*}
\big |\cG_{[1, \ell]}(z_0, \w, E)(j, k)\big | & \le \exp\Bigl(-{\gamma\over 2}(k - j) +K\Bigr)\\
\big \|\cG_{[1, \ell]}(z_0, \w, E)\big \| & \le \exp(K)
\end{align*}
where $\cG_{[1, \ell]}(z_0, \w, E_0) = \bigl(H_{[1,\ell]}(z_0, \w) -
E_0\bigr)^{-1}$ is the Green's function,  $1 \le j \le k \le \ell$.
\end{lemma}
\begin{proof} By Cramer's rule applied to $(H_{[1,\ell]}-E)^{-1}$, the uniform upper bound of
Lemma~\ref{lem:5.2}, as well as the rate of convergence estimate
$|L-L_\ell| \lesssim  \ell^{-1}$ from~\cite{GS1}, \beeq
\begin{aligned}
\big |\cG_{[1, \ell]}(z_0, \w, E)(j, k)\big | & = \big |f_{j-1}(z_0,
\w, E)\big | \cdot \big |f_{\ell-k}\bigl(z_0 e(k\w), \w,
E_0\bigr)\big | \cdot
\big |f_\ell(z_0, \w, E_0)\big|^{-1} \\
&\le \big |f_\ell(z_0, \w, E_0)\big |^{-1} \exp \bigl(\ell L(\w,
E_0) - (k - j) L(\w, E_0) + (\log\ell)^C\bigr),
\end{aligned}
\eneq and the lemma follows.
\end{proof}

The idea behind implementing the aforementioned Fr\"ohlich-Spencer
scheme is now as follows: First, if $\psi$ is an $\ell^2$-normalized
eigenfunction of $H_{[1,N]}(x_0,\omega)$ with eigenvalue~$E$,
then~\eqref{eq:large_det} must fail with $\ell = (\log N)^C$ for
some $z_0=e(x_0+k_0\omega)$. In other words, it must fail for some
determinant $f_{[k_0,k_0+\ell]}(x_0,\omega,E)$. Second, if it were
to fail for another determinant $f_{[k_1,k_1+\ell]}(x_0,\omega,E)$
where $|k_0-k_1|=t$ for a sufficiently large $t$, then this would
lead to a contradiction of the separation of zeros property
described above. Note that the latter dictates the size of $t$ (here
it turns out to be $t=\exp((\log\log N)^C)\;$) and therefore also
the size of the localization window. Moreover, note that we are
forced to eliminate a set of energies and $\omega$ to achieve this
separation of the zeros. Hence, we can only hope to obtain the
localization property of the eigenfunctions if the energy falls
outside a set of exceptional energies. We conclude that outside of
some window all determinants of a smaller scale
satisfy~\eqref{eq:large_det}, and thus the eigenfunction $\psi$ has
to be exponentially small there (due to an application of the
avalanche principle and~\eqref{eq:poisson}). For further details we
refer the reader to Section~10 of~\cite{GS2}.

\subsection{Distances between eigenvalues on a finite interval}

Dirichlet eigenvalues on a finite interval are simple. However, the
eigenvalues can be as close as $\exp(-cN)$, where $N$ is the size of
the interval. It follows for instance from the analysis of
(\ref{eq:1}) in Sinai's work~\cite{Si1} that this is the case if $x$
belongs to some subset of $\tor$ of measure $\exp(-\gamma N)$.  In
\cite{GS2} the localized eigenfunctions were used to improve upon
the $e^{-cN}$ bound. Indeed, the ``typical' distance between the
eigenvalues turns out to be~$e^{-N^\delta}$.

It will be convenient for us to work with the operators $H_{[-N,
N]}\xw$ instead of $H_{[1, N]}\xw$ as we did in the previous
section. We use the symbols $E_j^{(N)}, \psi_j^{(N)}$ to denote the
eigenvalues and normalized eigenfunctions of $H_{[-N, N]}\xw$. We
denote the sets of $\omega$ and energy, which we needed to remove in
the previous section, by $\Omega_N$, $\cE_{N,\w}$. They are of
measure $\lesssim \exp \bigl(-(\log N)^{A_2}\bigr)$ and complexity
$\lesssim \exp\bigl((\log N)^{A_1}\bigr)$ where $A_2 \gg A_1$.

\begin{lemma}
\label{lem:sep} For any $\w \in \tor_{c,a}\setminus \Omega_N$ and
all $x$ one has for all $j, k$ and any small $\delta > 0$ \beeq \big
|E_j^{(N)}\xw - E_k^{(N)}\xw \big | > e^{-N^\delta} \eneq provided
$E_j^{(N)}\xw \notin \cE_{N,\w}$ and $N \ge N_0(\delta)$.
\end{lemma}

\begin{proof} Fix $x \in \tor$, $E_j^{(N)}\xw\notin \cE_{N,\w}$.  Let $Q\asymp \exp\bigl((\log\log N)^C\bigr)$.
By the Anderson localization property (see the previous subsection
as well as Section~9 in~\cite{GS2}) there exists
$$
\Lambda_Q:= \left[ \nu_j^{(N)} \xw - Q,\ \nu_j^{(N)}\xw + Q\right] \cap \bigl[-N, N\bigr]
$$
so that \beeq\nn
\begin{aligned}
& \sum_{n \in [-N, N]\setminus \Lambda_Q} \big |f_{[-N, n]}\bigl(e(x), \w; E_j^{(N)}\xw\bigr)\big |^2\\
&\quad <e^{-2Q\gamma} \sum^N_{n=-N} \big| f_{[-N, n]}\bigl(e(x),\w;
E_j^{(N)}\xw\bigr)\big |^2\ .
\end{aligned}
\eneq
Here we used that with some $\mu = \const$
$$
\psi_j^{(N)} (x, \w; n) = \mu \cdot f_{[-N, n-1]}\bigl(e(x), \w; E_j^{(N)}\xw\bigr)
$$
for $-N \le n \le N$ and the convention that
$$
f_{[-N, -N-1]} = 0\ ,\quad f_{[-N, -N]} =1\ .
$$
One can assume $\nu_j^{(N)} \xw \ge 0$ by symmetry.  It follows from
the avalanche principle (recall that all determinants of scale
$(\log N)^C$ which fall outside of the window of localization are
non-resonant) and a simple stability bound in the energy that
\beeq\nn
\begin{aligned}
& \sum_{n = -N}^{\nu_j^{(N)}\xw - Q} \big | f_{[-N, n]}\bigl(e(x), \w; E\bigr) - f_{[-N, n]}\bigl(e(x), \w; E_j^{(N)}\xw\bigr)\big |^2\\
& \le e^{-2\gamma Q}\big |E - E_j^{(N)}\xw \big |^2 e^{(\log N)^C} \sum_{n \in \Lambda_Q}\big |f_{[-N, n]}\bigl(e(x),\omega; E_j^{(N)}\xw\bigr)\big|^2
\end{aligned}
\eneq Let $n_1 = \nu_j^{(N)} \xw - Q -1$.  Furthermore, we bound the
difference on the window of localization simply by
\beeq
\begin{aligned}
& \left\|\begin{pmatrix}
f_{[-N, n+1]}\bigl(e(x), \w, E\bigr)\\[5pt]
f_{[-N, n]}\bigl(e(x),\w, E\bigr)\end{pmatrix} -
\begin{pmatrix}
f_{[-N, n+1]}\bigl(e(x), \w, E_j^{(N)}\xw\bigr)\\[5pt]
    f_{[-N, n]}\bigl(e(x),\w, E_j^{(N)}\xw\bigr)\end{pmatrix}
\right\|\\
& = \left\|M_{[n_1+1,n]}\bigl(e(x), \w, E\bigr)
\begin{pmatrix}
 f_{[-N, n+1]}\bigl(e(x), \w, E\bigr)\\[5pt]
   f_{[-N, n]}\bigl(e(x),\w, E\bigr)\end{pmatrix} -
   M_{[n_1+1,n]}\bigl(e(x), \w, E_j^{(N)}\xw\bigr)
   \begin{pmatrix}
       f_{[-N, n+1]}\bigl(e(x), \w, E_j^{(N)}\bigr)\\[5pt]
             f_{[-N, n]}\bigl(e(x),\w, E_j^{(N)}\bigr)\end{pmatrix}
         \right\|\\
         & \le e^{C(n-n_1)} e^{-\gamma Q}
         \bigl|E - E_j^{(N)}\xw\big | e^{(\log N)^C}
         \Biggl(\sum_{n \in \Lambda_Q} \big |f_{[-N, n]}\bigl(e(x),\w, E_j^{(N)}\xw\bigr)\big |^2\Biggr)^{1/2}\ .
         \end{aligned}
       \nn  \eneq
Now suppose there is $E_k^{(N)}\xw$ with $\big
|E_k^{(N)}\bigl(e(x),\w\bigr) - E_j^{(N)}\xw\big | < e^{-N^\delta}$
for some small $\delta > 0$.  Then, by the preceding,
 \beeq
\begin{aligned}
& \sum^N_{n=-N} \big | f_{[-N, n]}\bigl(e(x), \w; E_j^{(N)}\xw\bigr) -  f_{[-N, n]}\bigl(e(x), \w; E_j^{(N)}\xw\bigr)\big |^2\\
& < e^{-{1\over 2}N^\delta} \sum_{n \in \Lambda_Q}\big |
f_{[-N, n]}\bigl(e(x), \w; E_j^{(N)}\xw\bigr)\big |^2\ ,
\end{aligned}
\nn\eneq provided $N^\delta > \exp \bigl((\log\log N)^A\bigr)$. That
contradicts the orthogonality of the eigenfunctions.
\end{proof}

\subsection{Simplicity of the zeros of $f_N(\cdot, \w, E)$}

An estimate for the minimal distance between the zeros of
$f_N(\cdot, \w, E)$ is crucial for  the analysis of the IDS
in~\cite{GS2}. In contrast with the eigenvalues of $H_N\xw$, $x \in
\tor$, the real zeros of the discriminant $f_N(\cdot, \w, E)$, $E
\in \IR$, can be degenerate. However, that happens only for special
values of the spectral parameter~$E$.  This follows from the
simplicity of the zeros of $f_N(x,\w, \cdot)$ by means of Sard-type
arguments. To turn this statement into a quantitative estimate one
has to make use of the estimate for the minimal distance between the
Dirichlet eigenvalues from the previous subsection.  The following
general assertion, which is a combination of Sard's theorem and
Cartan's estimate for analytic functions, allows one to do that.

\begin{lemma}
\label{lem:2weier} Let $f(z, w)$ be an analytic function defined in
$\cD(0,1)\times \cD(0,1)$. Assume that one has the following
representations:
\begin{enumerate}
\item[(i)] $f(z, w) = (w - b_0(z))\chi(z, w)$,
for any $z \in \cD(0, r_0)$, $w \in \cD(0, r_1)$, where $b_0(z)$ is
analytic in $\cD(0, r_0)$, $\sup|b_0(z)| \le 1$,  $\chi(z, w)$ is
analytic and non-vanishing on $\cD(0, r_0) \times \cD(0, r_1)$,
where $0<r_0,r_1<\frac12$

\item[(ii)] $f(z, w) = P(z, w)\theta(z, w)$, for any $z \in \cD(0, r_0)$,
$w \in \cD(0, r_1)$ where
$$
P(z, w) = z^k + c_{k-1}(w) z^{k-1} + \cdots + c_0(w)\ ,
$$
$c_j(w)$ are analytic in $\cD(0, r_0)$, and $\theta(z, w)$ is
analytic and non-vanishing on $\cD(0, r_0) \times \cD(0, r_1)$, and
all the zeros of $P(z, w)$ belong to $\cD(0, 1/2)$.
\end{enumerate}
Then given $H \gg k^2 \log [(r_0r_1)^{-1}]$ one can find a set
$\cS_H \subset \cD(w_0, r_1)$ with the property that
\[
\mes(\cS_H) \les r_1^2 \exp\left(-cH/k^2 \log
[(r_0r_1)^{-1}]\right), \text{\ \ and \ \ } \compl(S_H)\lesssim k^2
\log [(r_0r_1)^{-1}]
\]
such that for any $w \in \cD(0, r_1/2) \setminus \cS_H$ and $z \in
\cD(0, r_0)$ for which $w = b_0(z)$ one has
$$
\big | b'_0 (z) \big | > e^{-kH} 2^{-k} r_1\ .
$$
Moreover, for those $w$ the distance between any two zeros of
$P(\cdot,w)$ exceeds $e^{-H}$.
\end{lemma}
\begin{proof} Assume that $k\ge2$ and set $\psi(w) = \disc P(\cdot, w)$. If $k=1$, then skip to~\eqref{eq:11.6}.
Then $\Psi(w)$ is analytic in $\cD(0, r_1)$.  Assume that $|\psi(w)|
< \tau$ for some $\tau> 0$, $w\in \cD(0, r_1)$.  Recall that for any
$w$
\begin{equation}
\label{eq:11.2} \psi(w) = \prod_{i\ne j} \left(\zeta_i(w) -
\zeta_j(w)\right)\ ,
\end{equation}
where $\zeta_i(w)$, $i = 1, 2, \dots, k$ are the zeros of $P(\cdot,
w)$. Then $|\zeta_i(w) - \zeta_j(w)| < \tau^{2/k(k-1)}$ for some $i
\ne j$.  Set $\zeta_i = \zeta_i(w)$, $\zeta_j = \zeta_j(w)$.  Assume
first $\zeta_i \ne \zeta_j$.  Then
$$
f(\zeta_i, w) = 0\,\ f(\zeta_j, w) = 0,\quad 0 < |\zeta_i - \zeta_j|
< \tau^{2/k(k-1)}\ .
$$
Due to (i) one has $w = b_0(\zeta_i) = b_0(\zeta_j)$.  Hence,
\begin{equation}\label{eq:11.3}
|b'_0(\zeta_i)| \le {1\over 2} |\zeta_i - \zeta_j| \max |b''_0(z) |
\lesssim |\zeta_i - \zeta_j| r_0^{-2} < r_0^{-2} \tau ^{2/k(k-1)}\ .
\end{equation}
If $\zeta_i = \zeta_j$ then $P(\zeta_i, w) = 0$, $\partial_z
P(\zeta_i, w) = 0$.  Then $f(\zeta_i, w) = 0$, $\partial_z
f(\zeta_i, w) = 0$ due to the representation~(ii).  Then $w -
b_0(\zeta_i) = 0$, $b'(\zeta_i) = 0$ due to the representation~(i).
Thus (\ref{eq:11.3}) holds at any event. If $\varphi(z)$ is analytic
function in $\cD(0, r)$, then it follows from the general change of
variables formula that
$$
\mes \left\{w: w = \varphi(z),\ z \in \cD(0, r),\ |\varphi'(z)| <
\eta\right\} \le \pi r^2 \eta^2
$$
In view of the preceding one obtains
\begin{equation}
\label{eq:question} \mes \left\{w \in \cD(0, r_1): |\psi(w)| <
\tau\right\} \lesssim r_0^{-2} \tau^{2/k(k-1)}.
\end{equation}
On the other hand, due to \eqref{eq:11.2} one obtains
$$
\sup\left\{|\psi(w)|: w \in \cD(0, r_1)\right\} \le 1.
$$
Take $\tau \ll (r_0r_1)^{k(k-1)/2}$. Then one obtains from
(\ref{eq:question}) that
$$
|\psi(w) | \ge \tau
$$
for some $|w| < \frac{r_1}{2}$.  By Cartan's estimate there exists a
set $\cT_H \subset \cD\left(0, \frac{r_1}{2}\right)$ with
$$\mes \cT_H \lesssim r_1^2\exp\left(-cH/k^2 \log[
(r_0r_1)^{-1}]\right)
$$
and of complexity $\lesssim k^2 \log [(r_0r_1)^{-1}]$ such that
\begin{equation}
\label{eq:11.5} \log |\psi(w)| > -H
\end{equation}
for any $w \in \cD(0, {r_1\over 2}) \setminus \cT_H$.

In particular, (\ref{eq:11.5}) implies that \beeq \label{eq:zetaij}
|\zeta_i(w) - \zeta_j(w) | > e^{-H} \eneq for any $w \in \cD(0,
{r_1\over 2}) \setminus \cT_H$, $i \ne j$.  Take arbitrary $w_0$
such that $\dist(w_0, \cT_H) > 2e^{-H}$, $w_0 = b_0(z_0)$ for some
$z_0 \in \cD(0, r_0)$. Then
\[ |P(z,w_0)|\ge (2e^{H})^{-k} \text{\ \ for all\ \ }|z-z_0|= e^{-H}/2\]
by the separation of the zeros~\eqref{eq:zetaij}.   By our
assumption on the zeros of $P(z,w)$,
\[ \sup_{z\in\cD(0,r_0)}\sup_{w\in\cD(0,r_1)} |\partial_wP(z,w)| \lesssim r_1^{-1}.\]
Thus,
\[ |P(z,w)|>\frac12 2^{-k}e^{-kH} \text{\ \ if\ \ }|z-z_0|= e^{-H}/2,\quad |w-w_0|\ll 2^{-k}e^{-kH}r_1.\]
Then due to the Weierstrass preparation theorem,
\begin{equation}
\label{eq:11.6} P(z, w) = \bigl(z - \zeta(w)\bigr) \lambda (z, w)
\end{equation}
for any $z \in \cD(z_0, r'_0)$, $w \in \cD(w_0, r'_1)$, where $r'_0
= e^{-H}/2$, $r'_1 \ll e^{-kH} 2^{-k} r_1$, and  $\zeta(w)$ is an
analytic function in $\cD(w_0, r'_1)$, $\lambda(z, w)$ is analytic
and non-vanishing on $\cD(z_0, r'_0)\times \cD(w_0,r'_0)$.
Comparing the representation (i) and (\ref{eq:11.6}) one obtains
\begin{equation}
\label{eq:11.7}
\begin{cases}
w - b_0(z) = 0 & \text{iff}\\
z - \zeta(w) = 0
\end{cases}
\end{equation}
for any $z \in \cD(z_0, r'_0)$, $w \in \cD(w_0, r'_1)$.  It follows
from \eqref{eq:11.7} that
$$
\big | b'_0 \bigl(\zeta(w)\bigr)\big | \ge \big |\zeta'(w)\big
|^{-1} \gtrsim r'_1 \gtrsim e^{-kH} 2^{-k} r_1,
$$
as claimed.
\end{proof}

\subsection{Harnack's inequality and Jensen's formula for the logarithm of the norms of monodromy matrices}
\label{sec:harnack}

The logarithm of the norm of an analytic matrix-function is a
subharmonic function. Harnack's estimate in this context is not as
sharp as for the logarithm of the modulus of an analytic function.
The same comment applies to Jensen's averages. The latter here
refers to the following: the Jensen formula states that for any
function $f$ analytic on a neighborhood of $\cD(z_0,R)$,
see~\cite{levin}, \beeq \label{eq:jensen}
 \int_0^1 \log |f(z_0+Re(\theta))|\, d\theta - \log|f(z_0)|
 = \sum_{\zeta:f(\zeta)=0} \log\frac{R}{|\zeta-z_0|}
\eneq provided $f(z_0)\ne0$. We showed above how to combine this
fact with the large deviation theorem and the uniform upper bounds
to bound the number of zeros of $f_N$ which fall into small disks,
in both the $z$ and $E$ variables. In what follows, we will refine
this approach further. For this purpose, it will be convenient to
average over $z_0$ in~\eqref{eq:jensen}. Henceforth, we shall use
the notation
\begin{align}
\nu_f(z_0, r) &= \# \{z \in \cD(z_0, r): f(z)=0\} \label{eq:nudef}\\
J(u, z_0, r_1, r_2) &= \mathop{\nint}\limits_{\cD(z_0, r_1)} dx\, dy
\mathop{\nint}\limits_{\cD(z, r_2)} d \xi d \eta\, [u(\zeta)-u(z)].
\label{eq:Jdef}
\end{align}

The following simple lemma is proved in~\cite{GS2}. It is our main
tool for counting zeros.

\begin{lemma}
\label{lem:Jdef} Let $f(z)$ be analytic in $\cD(z_0, R_0)$. Then for
any $0<r_2<r_1<R_0-r_2$
\begin{equation*}
\nu_f(z_0, r_1 - r_2) \leq 4 \frac{r_1^2}{r_2^2} J(\log |f|, z_0,
r_1, r_2) \leq \nu_f(z_0, r_1+r_2)
\end{equation*}
\end{lemma}

We now describe how the aforementioned technical issues were
addressed in Section~12 of \cite{GS2} for the transfer matrices of
the Schr\"odinger co-cycles. More precisely, we state the two main
results of that section. The reader should not be distracted by
technicalities, but rather notice how the norms of the matrices
mimic the behavior of the entries. For the latter the crucial piece
of information is the number of zeros in various disks. In that
respect, we emphasize the quadratic estimate in~\eqref{eq:quad}. The
linear estimate (i.e., the one where the scalar logarithm is not
subtracted) would be too weak for the study of the IDS
in~\cite{GS2}.

\begin{prop}
\label{prop:5.10}
\begin{enumerate}
\item[{\rm{(i)}}] Suppose that one of the Dirichlet determinants
\[ f_{[1,
N]}(\cdot, \omega, E),\; f_{[1, N-1]}(\cdot, \omega, E),\; f_{[2,
N]}(\cdot, \omega, E),\; f_{[2, N-1]}(\cdot, \omega, E)
\]
has no zeros in
$\cD(z_0, r_1)$, $\exp(-\sqrt{N})\le r_1 \le \exp\bigl(-(\log N)^C\bigr)$.  Then
\end{enumerate}
\begin{equation}
\label{eq:quad} \Big | \log {\big \|M_N(z, \omega, E)\big \|\over
\big \| M_N(z_0, \omega, E)\big \|} - \log \big |1 + a_0(z -
z_0)\big | \Big | \le |z - z_0|^2 r_2^{-2}
\end{equation}
\begin{enumerate}
\item[]
for any $z \in \cD(z_0, r_2)$,  $r_2 = r_1\exp\bigl(-(\log
N)^{2C}\bigr)$, and with $|a_0| \lesssim r_2^{-1}$.

\item[{\rm{(ii)}}] Assume that the following conditions are valid
\begin{enumerate}
\item[{\rm{(a)}}] each of the determinants $f_{[a, N-b]}(\cdot, \omega,
E)$, $a = 1, 2$; $b = 0, 1$ has at least one zero in $\cD(\zeta_0,
\rho_0)$, where $e^{-\sqrt{N}}\le \rho_0 \le \exp\bigl(-(\log N)^{B_0}\bigr)$

\item[{\rm{(b)}}] no determinant $f_{[a, N-b]} (\cdot, \omega, E)$ has a
zero in $\cD(\zeta_0, \rho_1)\setminus \cD(\zeta_0, \rho_0)$, $\rho_1 \ge
\exp\bigl((\log N)^{B_1}\bigr)\rho_0$, $B_0 \gg B_1 +A$.
\end{enumerate}
Let $k_0 = \min\limits_{a, b} \cZ(f_{[a, N-b]} (\cdot, \omega, E),\zeta_0,\rho_0)$.
Then for any
\[ z, \zeta \in \cD(\zeta_0, \rho_1')\setminus \cD(\zeta_0,
\rho_2),\;\rho_1'=\exp\bigl(-(\log N)^{B_2}\bigr)\rho_1,\;\rho_2 = \exp\bigl((\log N)^{B_2}\bigr)\rho_0,\quad B_1 \gg
B_2\gg
1\]
 one has
$$
\Big | \log {\|M(\zeta)\|\over \|M(z)\|} - k_0 \log {|\zeta -
\zeta_0|\over |z - \zeta_0|} \Big | \le \exp\bigl(-(\log N)^{C}\bigr)
$$
\end{enumerate}
\end{prop}

\begin{prop}
\label{prop:5.11}
\begin{enumerate}
\item[{\rm{(i)}}] Assume that one of the Dirichlet determinants $f_{[a,
N-b]}(\cdot, \omega, E)$, $a = 1, 2$, $b = 0, 1$ has no zeros in
$\cD(z_0, r_1)$, $\exp(-\sqrt{N}) \le r_1 \le \exp\bigl(-(\log N)^{C_1}\bigr)$.  Then
\begin{equation}
\label{eq:15.9}
4{\rho_1^2\over \rho_2^2} J \left(\log \|M_N(\cdot, \omega, E)\|, z_0,
\rho_1, \rho_2\right) \le \rho_1^2 r_1^{-2} \exp\bigl((\log N)^B\bigr)
\end{equation}
for any $r_1\exp(-\sqrt{N}) \le \rho_1 \le r_1\exp\bigl(-(\log N)^A\bigr)$, $\rho_2 =c \rho_1$
\item[{\rm{(ii)}}] Assume that for some $\zeta_0$ the following
conditions are valid
\begin{enumerate}
\item[{\rm{(a)}}] each of the determinants $f_{[a, N-b]} (\cdot,
\omega, E)$, $a = 1, 2;$ $b = 0, 1$ has at least one zero in
$\cD(\zeta_0, \rho_0)$, $\exp(-\sqrt{N})<\rho_0 \le \exp\bigl(-(\log
N)^{B_0}\bigr)$.

\item[{\rm{(b)}}] no determinant $f_{[a, N-b]} (\cdot, \omega, E)$ has
a zero in $\cD(\zeta_0, \rho_1) \setminus \cD(\zeta_0, \rho_0)$, $\rho_1
\ge \exp\bigl((\log N)^{B_1}\bigr)\rho_0$, $B_0 > B_1$.
\end{enumerate}
\end{enumerate}
Let $k_0 = \min\limits_{a, b} \# \cZ(f_{[a, N-b]}(\cdot, \omega, E),\zeta_0,\rho_0)$.  Then for any
\[z_1 \in \cD(\zeta_0, \rho_1') \setminus \cD(\zeta_0, \rho_2),\;\rho_1'=\exp\bigl(-(\log N)^{B_2}\bigr)\rho_1,\;\rho_2
\asymp \exp\bigl((\log N)^{B_2}\bigr) \rho_0,\]
$B_1>B_2$, one has
$$
\Big | 4{r^2_1\over r_2^2} J\left(\log \|M_N(\cdot, \omega, E)\|, z_1,
r_1, r_2\right) - k_0 \Big | \le \exp \bigl(-(\log N)^C\bigr)
$$
where $|z_1-\zeta_0|(1+2c)<r_1 < \rho_1'$, $r_2 = c r_1$, and $0<c\ll 1$ is some constant.
\end{prop}

\section{The IDS: Lipschitz, H\"older, and absolute continuity}

We now sketch the main steps that allow us in~\cite{GS2} to pass
from information about the zeros of $f_N(z,\omega,E)$ in $z$ and $E$
to information on the IDS.

\subsection{Concatenation terms and the number of eigenvalues falling
into an interval}

Consider the following expressions which we call {\em concatenation
terms} in view of their role in the avalanche principle expansion:
\begin{equation}\label{eq:151}
\cW_{N, k}\bigl(e(x), E + i\eta\bigr) =
{\big \|M_{[1, k]}\bigl(e(x), \omega, E+i\eta\bigr)\big \|\, \big \|M_{[k+1, N]}\bigl(e(x), \omega, E + i\eta\bigr)\big \|\over
\big \|M_{[1, N]}\bigl(e(x), \omega, E+i\eta\bigr)\big \|}
\end{equation}
$1 \le k \le N$, where $\omega$ is fixed.

\begin{lemma}
\label{lem:151}
 Let $x \in \tor$, $E \in \IR$, $\eta > 0$, and let
\begin{equation}\label{eq:152}
\left | f_{[a, N-b+1]} \bigl(e(x), \omega, E+i\eta\bigr) \right | = \max_{1 \le a', b' \lesssim 2}
\left |f_{[a', N- b' +1]} \bigl(e(x), \omega, E+ i\eta\bigr)\right |
\end{equation}
for some $1 \le a$, $b \lesssim 2$.  Then
\begin{equation}
\label{eq:153} \#\left(\rsp H_{[a, N-b +1]} \bigl(e(x), \omega\bigr)
\cap \bigl(E - \eta, E+\eta\bigr) \right) \le 4 \eta \sum_{1 \le k
\le N}\, \cW_{N, k} \bigl(e(x), E + i\eta\bigr)
\end{equation}
\end{lemma}

\begin{proof} By Cramer's rule
\begin{align}
 & \left(H_{[a, N']} \bigl(e(x),\omega\bigr) - E - i\eta\right)^{-1}(k, k) =
{f_{[a, k]} \mape f_{[k+2, N']}\mape\over f_{[a, N']} \mape} \label{eq:154}\\[6pt]
& M_{[a, N']} \mape = \begin{bmatrix}
f_{[a, N']} \mape & - f_{[a+1, N']} \mape\\
f_{[a, N'-1]} \mape & - f_{[a+1, N'-1]}\mape
\end{bmatrix} \label{eq:155}
\end{align}
Due to (\ref{eq:152})
\begin{equation}
\label{eq:156} \big \|M_N \mape \big \| \le 2 \Big | f_{[a, N-b +1]}
\mape \Big |\ .
\end{equation}
Combining (\ref{eq:154}), (\ref{eq:155}), (\ref{eq:156}) one obtains
\begin{equation}
\label{eq:157}
\begin{aligned}
 & \left| \tr \left(\Bigl(H_{[a, N-b +1]} (x,\omega) -E - i\eta\Bigr)^{-1}\right)\right |\\
 & \le \sum_{a \le k \le N -b +1}\ {\Big |f_{[a, k]} \mape\Big |\, \Big |f_{[k+2, N-b +1]}\mape \Big |\over
 \Big |f_{[a, N-b +1]} \mape \Big |}\\
 & \le \sum_{a \le k \le N-b+1}\ 2 \cW_{N, k} (e(x),E+i\eta)
 \end{aligned}
 \end{equation}
 On the other hand,
 \begin{equation}\nn
 \begin{aligned}
 & \left| \tr \left(H_{[a, N-b +1]} (x, \omega) -E - i\eta\right)^{-1} \right|\\
 & \ge (2\eta)^{-1} \#\left(\rsp H_{[a, N-b+1]} \bigl(e(x), \omega\bigr) \cap (E - \eta, E+\eta)\right)
 \end{aligned}
 \end{equation}
 and we are done.
 \end{proof}

Recall that
\begin{equation}\label{eq:1511}
\begin{aligned}
\left | \left(H\mapen\right)^{-1} (k, k)\right| & \le \left \|
\left(H\mapen\right)^{-1}\right\| \le \eta^{-1}
\end{aligned}
\end{equation}

\begin{corollary}
\label{cor:5.13}
 Using the notations of Lemma \ref{lem:151} one has
\begin{equation*}
\begin{aligned}
& \# \left(\rsp \left(H_{[1, N]}\bigl(e(x),\omega\bigr)\right)\cap \bigl(E - \eta, E+\eta\bigr)\right)\\
& \le 4 \eta \sum_{ k \in K}\ \cW_{N,k}\bigl(e(x), E + i\eta\bigr) + \# (K) + 2
\end{aligned}
\end{equation*}
\end{corollary}

\begin{proof} Due to Weyl's Comparison Lemma, see~\cite{Bhat},
\begin{equation*}
\begin{aligned}
&\#\left(\rsp\left(H_{[1, N]}\bigl(e(x),\omega\bigr)\right) \cap \bigl(E - \eta, E+ \eta\bigr) \right) \le\\
& \#\left(\rsp\left(H_{[a, N-b+1]} \bigl(e(x), \omega \bigr) \right)\cap \bigl(E - \eta, E + \eta\bigr) \right) + 2
\end{aligned}
\end{equation*}
 Therefore, the assertion follows from the previous lemma.
\end{proof}

\begin{lemma}
\label{lem:5.14} Let $A$ be an $n\times n$ Hermitian matrix.  Let
$\Psi^{(1)}, \Psi^{(2)},\dots, \Psi^{(n)} \in \IC^n$ be an
orthonormal basis of eigenvectors of $A$ and $E^{(1)}, E^{(2)},
\dots, E^{(n)}$ be the corresponding eigenvalues.  Then for any $E +
i\eta$, $E \in \IR$, $\eta > 0$ one has
\begin{equation*}
\sum_{1 \le k \le n} \left| \left(\bigl(A - E - i\eta\bigr)^{-1} e_k, e_k \right)\right |^2  \ge \sum_{1 \le j \le n} \Biggl(\sum_{1 \le k \le n} \left |\bigl(e_k, \Psi^{(j)}\bigr) \right |^4 \Biggr) \cdot
\left(\imm \left(E^{(j)} - E - i\eta\right)^{-1}\right)^2
\end{equation*}
where $e_1, e_2, \dots, e_n$ is arbitrary orthonormal basis in $\IC^n$.
\end{lemma}

\begin{proof} One has
\begin{equation*}
\begin{aligned}
\left(\left(A - E - i\eta\right)^{-1} e_k, e_k \right) & = \sum_{1 \le j \le n} \left| \bigl(e_k, \Psi^{(j)}\bigr)\right |^2 \left(E^{(j)} - E - i \eta\right)^{-1}\\
\left|\left(\bigl(A - E - i\eta\bigr)^{-1} e_k, e_k \right) \right | & \ge \imm \left(\bigl(A - E - i\eta\bigr)^{-1} e_k, e_k \right)\\
& = \sum_{1 \le j \le n} \left|\bigl(e_k, \Psi^{(j)}\bigr) \right|^2 \imm \bigl(E^{(j)} - E - i\eta\bigr)^{-1}\ .
\end{aligned}
\end{equation*}
Since $\imm\left(E^{(j)} - E - i \eta\right)^{-1} > 0$, $j = 1,
2,\dots, n$, the assertion follows (use $(\sum_j a_j)^2\ge \sum_j
a_j^2$ if $a_j\ge0$).
\end{proof}
In the next corollary we show how to use effectively the localized
eigenfunction to sharpen the estimate on the number of eigenvalues
falling into an interval in an abstract setting.

\begin{corollary}
\label{cor:5.15}
Using the notations of the previous lemma assume that the following condition is valid for some $E, \eta$:

\bigskip\noindent
{\bf (L)\ } for each eigenvector $\Psi^{(j)}$ with $\left| E^{(j)} -
E\right| < \eta$ there exists a set $\cS(j) \subset
 \left\{1, 2, \dots, n\right\}$, $\#\cS(j) \le \ell$ such that $\sum\limits_{k \notin \cS(j)}
 \left|\bigl(e_k, \Psi^{(j)}\right)|^2 \le 1/2$.

\bigskip\noindent
 Then
$$
\#\left\{j: \left|E^{(j)} - E\right| < \eta\right\} \le 8\ell \eta^2
\sum_{1 \le k \le n} \left|\left(\bigl(A - E - i\eta\bigr)^{-1} e_k,
e_k \right) \right |^2
$$
\end{corollary}

\begin{proof} Recall that for any positive $\alpha_1, \dots, \alpha_\ell$ with $\sum\limits_j \alpha_j = 1$
\begin{equation}\nn
\sum_j \alpha^2_j \ge \sum_j\ {1\over \ell^2} = {1\over \ell}
\end{equation}
by Cauchy-Schwarz. Due to the assumptions of the corollary
$$
1 = \left(\Psi^{(j)}, \Psi^{(j)}\right) = \sum_{1 \le k \le n}
\left|\bigl(e_k, \Psi^{(j)})\right|^2 \le \sum_{k \in \cS(j)}
\left|\bigl(e_k, \Psi^{(j)}\bigr) \right|^2 + 1/2
$$
for any $\left|E^{(j)} - E\right | < \eta$.  Hence, for such
$E^{(j)}$ we have
\begin{equation}
\sum_{k \in \cS(j)} \left|(e_k, \Psi^{(j)})\right|^4 \ge 1/4\ell
\end{equation}
and the assertion follows from the previous lemma.
\end{proof}

Consider the following expressions which we also call concatenation
terms:
\begin{equation}\nn
w_m(z) = -\log\cW_m(z)= \log \frac{\big \|M_{2m}(z,\omega, E)\big\|
}{\big \|M_m \bigl(ze(m\omega), \omega, E\bigr) \big \|\, \big
\|M_m(z,\omega, E)\big \|}
\end{equation}
These terms are simple linear combinations of subharmonic functions.
Hence, the Riesz representation theorem allows one to write them in
the usual way, albeit with a signed measure rather than a positive
one. Nevertheless, it turns out that this measure is an "almost"
positive measure. This feature of the concatenation terms combined
with the avalanche principle expansion allows one to establish a
very sharp relation between these terms for different scales. Assume
that the following condition holds:

\bigskip
{\bf (I)} no determinant $f_{[a, m-b]}\bigl(\cdot e(nm\omega),
\omega, E\bigr)$, $f_{[a, 2m-b]}(\cdot, \omega, E)$, $a = 1, 2$; $b
= 0, 1$; $n = 0, 1$, has a zero in some annulus $\cD(\zeta_0,
\rho_1)\setminus \cD(\zeta_0, \rho_0)$, where $\rho_0 \asymp
\exp\bigl(-m^\delta\bigr)$, $\rho_0 < \rho_1 < \exp\left(-(\log
m)^A\right)$, $0 < \delta \ll 1$.
\bigskip

 Set
\begin{equation}\nn
\bar k_n = \min_{a, b}\, \nu_{f_{[a, m-b]}\bigl(\cdot
e(nm\omega),\omega,E\bigr)} (\zeta_0, \rho_0)\ ,
\end{equation}
$n = 0, 1$,
\begin{equation}\nn
\bar k = \min_{a, b}\, \nu_{f_{[a, 2m-b]}(\cdot, \omega,
E)}(\zeta_0, \rho_0)\ .
\end{equation}

\begin{lemma}\label{lem:16.1} Assume that $\rho_1 \ge \rho_0^{(\log m)^{-B_0}}$.  Then
\begin{equation}\nn
\bar k_0 + \bar k_1 \le \bar k \le \bar k_0 + \bar k_1 +
k_1(\lambda, V)
\end{equation}
provided $B_0 \gg 1$. Here $k_1(\lambda,V)$ is some integer
constant.
\end{lemma}

\begin{proof}
 Recall that due to the large deviation theorem there exists
 $z = e(x + iy) \in \cD(\zeta_0, \rho'_1) \setminus \cD(\zeta_0, \rho'_1/2)$ such that
\begin{equation}\nn
\left| \log \big \|M_{[1, m]}\bigl(ze(nm\omega),\omega, E\bigr) \big
\| - m L(y, E) \right| \le m^\delta\bigl(\log m\bigr)^{-B_2}\ ,\ n =
0, 1,
\end{equation}
\begin{equation}\nn
\left | \log \big \|M_{[1, 2m]}(z, \omega, E) \big \| - 2m L(y, E)
\right| \le m^\delta \bigl(\log m\bigr)^{-B_2}
\end{equation}
with $1\ll B_2 < B_0$. Combining these relations with
Proposition~\ref{prop:5.10} one obtains
\begin{equation}\nn
\begin{aligned}
& \biggl | \log {\big \|M_{[1, m]} \bigl(\zeta e(m\omega), \omega, E\bigr) \big\|\, \big \|M_m(\zeta, \omega, E)\big \|\over
\big \|M_{[1,2m]}(\zeta, \omega, E)\big \|} -\\
&\quad \left(\bar k_0 + \bar k_1 - \bar k\right) \log {|\zeta - \zeta_0|\over |z - \zeta_0|} \bigg| \le
Cm^\delta\bigl(\log m\bigr)^{B_1}
\end{aligned}
\end{equation}
for any $\zeta \in \cD(\zeta_0, \rho'_1) \setminus \cD(\zeta_0,
\rho_2)$. Since $|z - \zeta_0| \asymp \rho'_1$, $\rho'_1 \asymp
\exp\left(-m^\delta (\log m)^{-B_0} - (\log m)^{B_1}\right)$, one
can pick $\zeta \in \cD(\zeta_0, \rho'_1) \setminus \cD(\zeta_0,
\rho_2)$ such that $|\zeta - \zeta_0| = |z - \zeta_0|^{(\log
m)^{B_0/2}}$.  Then
\begin{equation}\label{eq:16.10}
\begin{aligned}
& \bigg | \log {\big \|M_{[1, m]} \bigl(\zeta e(m\omega), \omega, E\bigr) \big \|\, \big \| M_m(\zeta, \omega, E)\big \|\over
\big \|M_{2m} (\zeta, \omega, E)\big \|} +\\
& m^\delta (\log m)^{B_0/2} \left(\bar k_0 + \bar k_1 - \bar k\right) \bigg | \le
Cm^\delta (\log m)^{B_1}\ .
\end{aligned}
\end{equation}
Recall that
\begin{equation}\label{eq:16.11}
\big \|M_{2m} (\zeta, \omega, E) \big \| \le \big \|M_{m}\bigl(\zeta e(m\omega),
\omega, E\bigr) \big \|\, \big \|M_m (\zeta, \omega, E)\big \|
\end{equation}

Relations (\ref{eq:16.10}), (\ref{eq:16.11}) imply $\bar k_0 + \bar
k_1 - \bar k \le 0$.  Removing the absolute values
in~\eqref{eq:16.10} and~\eqref{eq:16.11} and taking Jensen's
averages one obtains the following:
\begin{equation}\nn
\nn \bigg | 4\, {r_2^2\over r_1^2}\ J\biggl (\log {\big
\|M_{2m}(\cdot, \omega, E)\big \|\over \big \|M_{[1, m]}\bigl(\cdot
e(m\omega), \omega, E\bigr)\big \|\, \big \|M_m(\cdot, \omega,
E)\big \|}\ ,\ \zeta_0, r_1, r_2\biggr ) - \left(\bar k - \bar k_1 -
\bar k_2\right)\bigg | < 2
\end{equation}
where $r_1=\rho_1'/2$, $r_2=cr_1$.  Hence, $\bar k - \bar k_1 - \bar
k_2 \le k_1(\lambda, V)$.
\end{proof}

The integers defined in the previous lemma almost perfectly
substitute the measure representing the concatenation term. More
precisely,  the following assertion holds:

\begin{prop} \label{prop:165}
Given $E \in \IC$ and integer $m \gg 1$, there exists a cover of
$\cA_{\rho_0/2}$ by disks $\cD\left(\zeta_{j, m}, \bar\rho_{j, m}\right)$,
$\zeta_{j, m} \in \cA_{\rho_0/2}$, $j = 1, 2,\dots, j_m$ such that the following conditions are valid:
\begin{enumerate}
\item[(1)] $\exp(-m^\delta) \le \bar \rho_{j, m} \le \exp(-m^{\delta/2})$, $j = 1, 2,\dots, j_m$, $0 < \delta \ll 1$,

\item[(2)] $\dist\left(\cD\left(\zeta_{j_1, m}, \underline \rho_{j_1, m}\right), \cD\left(\zeta_{j_2, m},\underline\rho_{j_2, m}\right)\right) \ge \underline\rho_{j_1, m} + \underline\rho_{j_2, m}$ where $\underline\rho_{j, m} = \bar\rho_{j, m}^{(\log m)^{B_1}}$, $B_1 \gg 1$, $j = 1, 2,\dots, j_m$, provided $j_1 \ne j_2$,

\item[(3)] no determinant $f_{[a, m-b]}\left(\cdot e(nm\omega), \omega, E\right)$ or $f_{[a, 2m-b]}(\cdot, \omega, E)$, $a = 1, 2$; $b = 0, 1$; $n = 0$, has a zero in $\cD\left(\zeta_{j, m}, \bar{\bar\rho}_{j, m}\right) \setminus \cD\left(\zeta_{j, m}, \underline{\underline\rho}_{j, m}\right)$, where $\underline{\underline\rho}_{j, m} = \underline\rho_{j, m}^{(\log m)^{B_1}}$, $\bar{\bar\rho}_{j, m} = \bar\rho_{j, m}^{(\log m)^{-B_1}}$, $j = 1, 2,\dots, j_m$,

\item[(4)] for each $\zeta_{j, m}$ there is an integer $k(j, m)$,
\begin{equation}\nn
0 \le k(j, m) \le \nu_{f_{2m}(\cdot, \omega, E)}\left(\zeta_{j, m}, \underline\rho_{j, m}\right)
\end{equation}
such that for any $z, \zeta \in \cD\left(\zeta_{j, m}, 2\bar\rho_{j, m}\right)\setminus \cD\left(\zeta_{j, m}, \underline\rho_{j, m}/2\right)$ holds
\begin{equation}\nn
\left| \left(w_m(\zeta) - w_m(z)\right) - k(j, m) \log {|\zeta -
\zeta_{j, m}|\over |z - \zeta_{j, m}|} \right| \le |\zeta - z|^2
\cdot \left(\bar\rho_{j, m}\right)^{-2}\ .
\end{equation}
\end{enumerate}
\end{prop}

\noindent Assume that the following condition is valid:

\bigskip \noindent{\bf (II.m)}  no determinant $f_{[a,
m-b]}\bigl(\cdot e(nm\omega), \omega, E\bigr), f_{[a, 2m-b]}(\cdot,
\omega, E)$ has more than one zero in any disk \[\cD\bigl(z_0,
r_m\bigr),\; r_m = \exp\left(-(\log m)^A\right),\; z_0 \in
\cA_{\rho_0/2},\; n=0,1,\; a=1,2,\; b=0,1 \]

\medskip

Consider two concatenation terms $w_m(z)$ and $w_\ulm(z)$
 with $m \asymp \exp
\left(\ulm^{\delta_1}\right)$, $0 < \delta_1 \ll 1$. Let
$\cD\left(\zeta_{j,m}, \bar\rho_{j, m}\right)$, $j = 1, 2,\dots,
j_m$ and $\cD\left(\zeta_{j, \ulm}, \bar\rho_{j, \ulm}\right)$, $j =
1, 2,\dots, j_{\ulm}$ be the disks defined in
Proposition~\ref{prop:165} for $w_m(z)$ and $w_\ulm (z)$,
respectively. Note that due to the avalanche principle expansion we
can conclude the following:

\begin{lemma}\label{lem:171} There exists $\cF_{\ulm, \omega, E} \subset \capo$ with
$\mes \cF_{\ulm, \omega, E} \le \exp\bigl(-\ulm^{1/2}\bigr)$ such that
\begin{equation}\nn
\Bigm |w_m(z) - w_\ulm\left(ze(m - \ulm)\omega\right)\Bigm | < \exp \left(-\ulm^{1/2}\right)
\end{equation}
for any $z \in \cA_{\rho_0/2} \setminus \cF_{\ulm, \omega, E}$.
\end{lemma}

Combining this assertion with the preceding one obtains the
following Proposition (using the notations of
Proposition~\ref{prop:165}).

\begin{prop}
Assume that conditions (II.$\ulm$), (II.m) are valid.  Then, using
the notations of the previous lemma one has
\begin{enumerate}
\item[(0)] if $k(j_1, \ulm) = 0$, then there exists $\zeta_{j, m} \in \cD\left(\underline\zeta_{j_1, \ulm},
\underline\rho_{j_1, \ulm}\right)$ with $k(j, m) = 0$ such that
$$
\left |w_m(z) - w_\ulm \left(z_1 e\bigl((m - \ulm)\omega\bigr)\right)\right | \lesssim \exp\left(-\ulm^{1/2}\right)
$$
for any $z \in \cD\left(\underline\zeta_{j_1, \ulm}, \bar\rho_{j_1,
\ulm}\right)\setminus \cD\left(\zeta_{j, m}, \underline\rho_{j,
m}\right)$ and any \[ z_1 \in \cD\left(\zeta_{j_1, \ulm},
\bar\rho_{j_1, \ulm}\right)\setminus \cF_{\ulm, \omega, E},\quad
\mes\cF_{\ulm, \omega, E} < \exp\left(-\ulm^{1/2}\right)\]

\item[(1)] if $k(j_1, \ulm) = 1$, then there exists $\zeta_{j, m} \in \cD\left(\underline\zeta_{j_1, \ulm}, \bar\rho_{j_1, \ulm}\right)$ with $k(j, m) = 1$ such that
$$
\left |w_m(z) - w_\ulm \left(\zeta e\bigl((m -\ulm)\omega\bigr)\right) - \log \left(\tfrac{|z - \zeta_{j, m}|}{|\zeta e\bigl((m - \ulm)\omega\bigr) - \zeta_{j, m}|}\right)\right|
< \exp\left(-\ulm^{1/2}\right)\ ,
$$
for any $z \in \cD\left(\underline\zeta_{j_1, \ulm}, \bar\rho_{j_1, \ulm}\right)\setminus \cD\left(\zeta_{j, m},
\underline\rho_{j, m}\right)$, $\zeta \in \cD\left(\underline\zeta_{j_1, \ulm},
\bar\rho_{j_1, \ulm}\right)\setminus \cF_{\ulm, \omega, E}$.
\end{enumerate}
\end{prop}

One can iterate the estimates of the previous proposition over a
decreasing series of scales. In fact, this can be done all the way
down to a scale of unit size. In this fashion one arrives at the
following:

\begin{theorem}
\label{thm:2} Let $V(x)$ be real analytic.  Assume $L(\omega_0, E)
\ge \gamma_0 > 0$ for some $\omega_0 \in \tor_{c,a}$ and all $E \in
(E', E'')$ and fix $b>0$ small.  There exist $N_0 = N_0(\lambda , V,
\gamma_0,b,c,a)$ , $\tau_0 = \tau_0(\lambda, V, \gamma_0,b,c,a) > 0$
so that:

For any $\vep > 0$, there exists $\Omega(\vep) \subset \tor$, $\mes
\Omega(\vep) < \vep$ such that for any $\omega \in (\omega_0 -
\tau_0, \omega_0 + \tau_0) \cap \left(\tor_{c,a} \setminus
\Omega(\vep)\right)$, there exists $\cE_\omega(\vep) \subset \IR$,
$\mes \cE_\omega(\vep) < \vep$ such that for any $N > N_0$ and any
$E \in (E', E'') \setminus \cE_\omega(\vep)$ and any $\eta >
1/N(\log N)^{1+b}$, one has
\begin{equation}
\label{eq:1.9} \int_\tor \#\left(\rsp\left(H_N(x, \omega)\right)\cap
\left(E - \eta, E+\eta\right)\right)\,dx \le \exp\left((\log
\vep^{-1})^A\right)\eta N\ .
\end{equation}
In particular, the IDS satisfies
\begin{equation}\nn
 \cN(E + \eta) - \cN(E - \eta) \le \exp\left((\log
\vep^{-1})^A\right)\eta
\end{equation}
for any $E \in (E', E'') \setminus \cE_\omega(\vep)$, $\eta > 0$.
\end{theorem}

The proof of Theorem~\ref{thm:2} establishes the estimate
\eqref{eq:1.9} for any $E \in \IR \setminus \cE_{\omega}(\vep)$,
with very detailed description of $\cE_{\omega}(\vep)$ as a union of
intervals of different scales. This allows one to combine the
Lipschitz estimate here with the H\"older bound of
 Theorem~\ref{thm:1} below to prove
the following

\begin{theorem}
\label{thm:3} For almost all $\omega \in (\omega_0 - \tau_0,
\omega_0 + \tau_0)$ the IDS $\cN(E)$ is absolutely continuous on
$(E', E'')$.  In particular, if $L(\omega_0, E) \ge \gamma_0 > 0$
for all $E$, then $\cN(\cdot)$ is absolutely continuous everywhere.
\end{theorem}

\begin{proof}[Proof of Theorem~\ref{thm:3}]  Let $\left(E'_n, E''_n\right)$,
$1 \le n \le \bar n$ be disjoint intervals with $\varepsilon =
\sum\limits_n \left(E''_n - E'_n\right) \ll 1$. Set $\tau =
\min\limits_n \left(E''_n - E'_n\right)$.  Let $\omega_r = p_r
q_r^{-1}$ be a convergent of $\omega$ with $q_r > \tau^{-4}$.  Let
$m^{(s)}$, $s = 1, 2,\dots, t+1$ be integers such that: (1)~ $\log
\bigl( m^{(s+1)}\bigr) \asymp \big(m^{(s)}\big)^\delta$, $s = 1,
2,\dots,t$, (2)~$\varepsilon >
\exp\left(-m^{(1)}\right)>\sqrt{\varepsilon}$, $m^{(t+1)} = q_r=:N$.
Using the Lipschitz estimates of the previous theorem applied to
each pair of consecutive scales $m^{(s)}$ , $m^{(s+1)}$ one can
obtains
$$
{1\over N} \int \#\Bigl(\rsp H_N(x,\omega)\cap \Bigl({\ell\over
N^{1/2}}\,  {\ell+1\over N^{1/2}}\Bigr)\Bigr)\, dx \lesssim m^{(1)}
N^{-1/2}
$$
for any interval $\left({\ell\over N^{1/2}}\ , {\ell +1\over
N^{1/2}}\right)\subset \IR \setminus \bigcup\limits^{t+1}_{s=1}\,
\cE_\omega^{(s)}$, where  $\ell \in \IZ$, provided $\omega \in
\tor_{c,a} \setminus \bigcup\limits_s \Omega^{(s)}$. Let
$\left\{T_\ell: \ell \in \cL\right\}$ be the collection of such
intervals.  Then
\begin{equation}\nn
{1\over N} \int \#\left(\rsp H_N(x,\omega) \cap \Biggl(\bigcup_{\ell \in \cL, T_\ell
\subset \bigcup\bigl(E'_n, E''_n\bigr)} T_\ell\Biggr)\right) dx \le m^{(1)} \varepsilon\ .
\end{equation}
Let $\cL' = \left\{\ell \in \cL: T_\ell\cap \left\{E'_n,
E''_n\::\: n = 1, 2,\dots, \bar n\right\} \ne \emptyset \right\}$.
Then $\#\cL' \le 2\bar n$.  Since $\bar n \lesssim \tau^{-1}$ one
obtains:
\begin{equation}\nn
{1\over N} \int \#\left(\rsp H_N(x,\omega)\cap \Biggl(\bigcup_{\ell
\in \cL'} T_\ell\Biggr)\right) dx \lesssim m^{(1)} N^{-1/2}\bar n
< m^{(1)} \tau\ .
\end{equation}
Finally, using the H\"older bound of Theorem~\ref{thm:1} as well as
the measure and complexity bounds on the exceptional sets
$\cE^{(s)}$ yields
$$
{1\over N} \int \#\left(\rsp H_N(x,\omega)\cap
\Biggl(\bigcup^{t+1}_{s=1} \cE^{(s)}_\omega\Biggr)\right) dx <
\exp\left(-(\log m^{(1)})^A\right)
$$
and we are done.
\end{proof}

Note that the previous proof exploits detailed information on the
size and complexity of those sets on which the IDS is not Lipschitz
(the {\em exceptional set}). This is necessary, as can be seen from
the example of a Cantor staircase function. Indeed, in that case
there is a uniform H\"older bound with an exponent that equals the
Hausdorff dimension of the Cantor set. However, in our case the
exceptional set has Hausdorff dimension zero, whereas the H\"older
exponent is fixed and positive.

\subsection{The H\"older bound}

The following result is proved in \cite{GS2}.

\begin{theorem}
\label{thm:1} Let $V_0\bigl(e(x)\bigr) = \sum\limits^{k_0}_{-k_0}
v(k) e(kx)$ be a trigonometric polynomial, $v(-k) =
\overline{v(k)}$, $-k_0 \le k \le k_0$. Let $L(E, \omega_0)$ be the
Lyapunov exponent  for $V = V_0$ and some $\omega_0 \in \tor_{c,a}$.
Assume that it exceeds $\gamma_0$ for all $E \in (E',E'')$.
\begin{enumerate}
\item[{\rm{(1)}}]
Given $\rho_0 > 0$ there exists $\tau_0 = \tau_0
(\lambda,V_0,\omega_0,\gamma_0,\rho_0)$ with the following property:
 for any 1-periodic, analytic function $V(e(x+iy))$, $- \rho_0 <
y < \rho_0 $ assuming real values when $y=0$ and  deviating from
$V_0(e(x))$ by at most $\tau_0$,
 any $\omega
\in \tor_{c,a} \cap (\omega_0 - \tau_0,\omega_0 + \tau_0) $, and any
$E \in (E',E'')$, with $\eta = N^{-1+\delta}$, $\delta \ll 1$, $N
\gg 1$, one has
\end{enumerate}
\begin{equation}
\int_{\tor} \#\left(\rsp\bigl(H_N(x, \omega) \bigr) \cap \bigl(E -
\eta, E + \eta\bigr)\right) dx   \le \eta^{\frac{1}{2k_0}-\eps}
\cdot N \label{eq:1.8}
\end{equation}
\begin{enumerate}
\item[]
with some constant $1 \ll B$ and arbitrary $\eps>0$.

\item[{\rm{(2)}}] The IDS  $\cN(\cdot)$ satisfies, for any small
$\eps>0$,
$$
\cN(E+ \eta) - \cN(E - \eta) \le \eta^{\frac{1}{2k_0}-\eps}\ ,
$$
for all $E\in (E',E'')$ and all small $\eta>0$.
\end{enumerate}
\end{theorem}

For the case of the almost Mathieu equation~\eqref{eq:1} (which
corresponds to $k_0=1$) and large $\lambda$, Bourgain~\cite{Bou1}
had previously obtained a H\"older-$(\frac12-\eps)$ result for the
IDS, which is known to be optimal in those regimes, see~\cite{Si1}.
See also~\cite{Bou2}.

The proof of Theorem~\ref{thm:1} is similar to that of
Theorem~\ref{thm:2} above. Recall that the latter result exploited
the fact that as long as we remove all energies $E$  belonging to
some bad set $\cE_\omega$, the zeros of $f_N(z,\omega,E)$ in~$z$ do
not cluster. In fact, any small disk (say of size $e^{-N^\delta}$)
does not contain more than one zero.

The logic is that here we can no longer guarantee this separation
property of the zeros since we are not allowed to remove energies.
Nevertheless, we will be able to show that zeros cannot cluster too
much in any small disk. The argument proceeds by contradiction: If
there were too many (in fact, $2\deg(V)+1$ many) zeros in a small
disk, then we can show that this would have to be the case in a
large number of disks of the same size (by exploiting the dynamics).
Ultimately, this leads to a contradiction due to the fact that the
determinant $f_N$ cannot have more than $2N\deg(V)$ zeros in total.
This fact appears to be of independent interest, and is formulated
as a theorem in~\cite{GS2}:

\begin{theorem}
\label{thm:4} Using the notations of Theorem \ref{thm:1} there
exists $k_0(\lambda, V) \le 2 \degg V_0$ with the following
property: for all $E \in \IR$, $s\in\ZZ$
 and $\omega \in
\tor_{c,a}$ and any $x_0 \in \tor$ there exists $s^{-}, s^{+}$ with
$|s-s^{\pm}| < \exp\bigl((\log s)^{\delta}\bigr)$ such that the
Dirichlet determinant $f_{[-s^-, s^+]}(\cdot, \omega, E)$ has no
more that $k_0(\lambda, V)$ zeros in $\cD\bigl(e(x_0), r_0\bigr)$,
$r_0 \asymp \exp\left(-(\log s)^A\right)$.
\end{theorem}

While we need to refer the reader to \cite{GS2} for more details, we
do present some basic statements here, which elucidate the role of
Jensen averages and the avalanche principle in this context.

\begin{defi}
\label{def:adj}
 Let $\ell\gg 1$ be some integer, and $s\in\ZZ$. We say that $s$ is {\em adjusted }
to a disk $\cD(z_0,r_0)$ at scale $\ell$ if for all $k\asymp\ell$
\[ \cZ(f_{k}(\cdot e((s+m)\omega),\omega,E),z_0,r_0)=\emptyset \qquad \forall\;|m|\le C\ell. \]
\end{defi}

Consider the avalanche
  principle expansion of $\log \big | f_N(z, \omega, E)\bigr |$:
  \begin{equation}
\label{eq:15.25}
  \log \big | f_N(z, \omega, E+i\eta)\big | = \sum^{n-1}_{m=1} \log
  \big \|A_{m+1}(z) A_m(z) \big \| - \sum^{n-1}_{m=2} \log \big \| A_m(z)\big \| +
  O\left(\exp\bigl(-\ell^{1/2}\bigr)\right)\ ,
  \end{equation}
  for any $z \in \cA_{\rho_0/2}\setminus \cB_{E,\eta,\omega}$, $\mes
  \cB_{E,\eta, \omega} \le \exp\bigl(-\ell^{1/2}\bigr)$, where $A_m(z)
  = M_\ell\bigl(ze(s_m\omega), \omega, E+i\eta\bigr)$, $m = 2,\dots,
  n -1$, $A_1(z) = M_{\ell_1}(z, \omega, E) \begin{bmatrix} 1 & 0\\ 0
  & 0\end{bmatrix}$, $A_n(z) = \begin{bmatrix} 1 & 0\\ 0 &
  0\end{bmatrix}M_{\ell_n}\bigl(ze(s_n\omega), \omega, E\bigr)$,
  $\ell_m = \ell$, $m = 1, 2,\dots, n-1$, $\ell_n = \tilde\ell$,
  $(n-1)\ell + \tilde\ell = N$, $\ell, \tilde \ell \asymp (\log N)^A$,
  $s_m = \sum\limits_{j< m} \ell_j$.

This expansion allows us to control the number of zeros of the large
scale object (in this case $f_N$) by means of the number of zeros
(or rather, the Jensen averages) of the small-scale objects (here
$w_j$) and vice versa. Surprisingly, it turns out that the most
effective way to implement this idea is to obtain an estimate for
the local number of zeros at a smaller scale in terms of the total
number of zeros at a larger scale. The all-important quadratic (more
precisely, super-linear) error estimate here is due
to~\eqref{eq:quad} above.

\begin{lemma}
\label{lem:15.8}
Assume that
$\{s_{m_j}\}_{j=1}^{j_0}$ is adjusted to
$\cD(z_0, r_0)$ at scale $\ell$. Set $m_0=0$, $m_{j_0+1}=n$, and
\[
w_j(z)   = \log \Big \| \prod^{m_j+1}_{m = m_{j+1}} A_m(z)\Big\| \text{\ \ for any\ \ }0\le j\le j_0
\]
Then
\begin{equation}
 4{r_1^2\over r_2^2} \Big | J\left(\log \big |f_N(\cdot, \omega, E)\big
|, z_0, r_1, r_2\right)  - \sum_{j=0}^{j_0} J(w_j(\cdot), z_0, r_1, r_2)
\Big |
\le N\exp\bigl((\log \ell)^C\bigr)\, r_1^2 r_0^{-2}
\label{eq:15.29}
\end{equation}
for any $e^{-\sqrt{\ell}}<r_1\les\exp(-(\log \ell)^A)r_0$, and $r_2=cr_1$.
In particular,
\begin{equation}
 4 {r_1^2\over r_2^2}\ J\left(\log \big | f_N(\cdot, \omega, E)\big |,
z_0, r_1, r_2\right)  \ge \sum_{j\in\cJ}\ J\bigl(w_j(\cdot), z_0, r_1,
r_2\bigr) -N\exp\bigl((\log
\ell)^C\bigr)\,r_1^2 r_0^{-2}
\end{equation}
for any $\cJ\subset [0,j_0]$.
\end{lemma}

For the remaining details (in particular, the crucial notion of
"contributing" terms) we refer the reader to~\cite{GS2}.

\section{Generic $C^3$ potentials}

In this section we review some recent work of Jackson Chan,
see~\cite{Cha}.  Given any function $V: \tor \rightarrow \IR$, we
have a family of quasi-periodic discrete Schr\"{o}dinger equations
\begin{equation}\label{eq:7.1}
-\varphi(n+1)-\varphi(n-1)+\lambda V(x+n\w)\varphi(n)=E\varphi(n), \qquad n \in \mathbb{Z}
\end{equation}
where $(x,\w) \in \tor \times \tor$ are parameters. Equation
(\ref{eq:7.1}) can be rewritten as a first order difference
equation:
$$
\left( \begin{array}{c} \varphi(n+1) \\ \varphi(n) \end{array}
\right) = \left( \begin{array}{cc} \lambda V(x+n\w)-E & -1 \\ 1 & 0
\end{array} \right) \left( \begin{array}{c} \varphi(n) \\ \varphi(n-1)
\end{array} \right).
$$
Given a $C^3$ potential $V$, any $C^3$ function $\tilde V$
satisfying the conditions
\begin{align*}
\max_{x\in\tor} | V(x) - \tilde V(x) | & < \delta \\
\max_{x\in\tor} | V'(x) - \tilde V'(x) | & < \delta \\
\max_{x\in\tor} | V''(x) - \tilde V''(x) | & < \delta
\end{align*}
can be written, near $x=0$, in the form
$$
\tilde V(x) = V(x) + \eta + \xi x + {1\over 2} \theta x^2 + x^3 R(x)
$$
where $|\eta|, |\xi|, |\theta| < \delta$, $R \in C^3$,
$|\partial_\alpha R| \lesssim 1$ for any index $ |\alpha| \le 2$.
More generally, since $\tor$ is compact, we can find some large
integer $T$ so that
\begin{equation}
\tilde V(x) = V(x) + \sum\limits_{m=1}^{T} \Bigl[ \eta_m + \xi_m \bigl(
x-{m\over T} \bigr) + {1\over 2} \theta_m \bigl( x-{m\over T} \bigr)^2 +
\bigl( x- {m\over T} \bigr)^3 R_m \bigl( x-{m\over T} \bigr) \Bigr]
\end{equation}
for all $x \in \tor$, where $\eta=(\eta_1, \ldots , \eta_T),\
\xi=(\xi_1, \ldots , \xi_T),\ \theta=(\theta_1, \ldots , \theta_T)
\in \prod\limits_1^T [-\delta,\delta]$, and $R_m \in C^3$,
$|\partial_\alpha R_m| \lesssim 1$ for any index $ |\alpha| \le 2$.
This motivates the following definition.

\begin{defi}
Let $T$ be a large integer, $0< \delta \ll {1\over T^5}$.
Suppose $R_m(\eta_m,\xi_m,\theta_m;x)$ are $C^3$ functions,
$m=1,2, \ldots, T$, $(\eta,\xi,\theta) \in \prod\limits_1^{3T} [-\delta,\delta]$,
$x \in \tor$, satisfying the following conditions:
\end{defi}
\begin{align}\nn
|\partial_\alpha R_m(\eta_m,\xi_m,\theta_m;x)| \lesssim  {1\over T} &
\qquad \text{for any index $|\alpha|\le 3$} \\
R_m(0,0,0;x) \equiv 0 & \nn\\
R_m(\eta_m,\xi_m,\theta_m;x)=-x^{-3}\bigl( \eta_m + \xi_m x +
{1\over 2}\theta_m x^2 \bigr)& \qquad \text{for $|x| \ge {1\over
2T}$}\nn
\end{align}
Define a $(T,\delta)$--variation of the potential by
$$
W(\eta,\xi,\theta,\{ R_m \} ; x) = \sum\limits_{m=1}^{T}
v_m\Bigl(\eta_m,\xi_m,\theta_m;x-{m\over T} \Bigr)
$$
where
$$ v_m(\eta_m,\xi_m,\theta_m;x) = \eta_m + \xi_m x + {1\over 2} \theta_m x^2 +
x^3 R_m (\eta_m,\xi_m,\theta_m;x) $$ By the preceding,
\begin{align*}
v_m(0, 0, 0; x) & \equiv 0 \\
\intertext{and}
v_m(\eta_m, \xi_m, \theta_m; x) & = 0\qquad \text{for $|x| \ge {1\over 2T}$}\ .
\end{align*}
Denote the collection of $(T,\delta)$--variations of the potential
by $\cS(T,\delta)$. The set of parameters $(\eta, \xi, \theta)$ has
measure $(2 \delta)^{3T}$. We want to define a notion of ``typical''
potential by using the normalized measure on this set of parameters.
Hence, a set $S \subset \cS(T,\delta)$ is called
$(1-\varepsilon)$-typical if
$$
|S| := \min\limits_{ \{ R_m \} } {1\over (2\delta)^{3T}}
\mes\bigl\{(\eta, \xi, \theta) \in [-\delta,\delta]^{3T}:
W(\eta, \xi, \theta, \{ R_m \} ; . ) \in S \bigr\} \ge 1 - \varepsilon
$$

\begin{theorem}
\label{thm:chan} Given any $V \in C^3(\tor)$, there is
$\lambda_0=\lambda_0(V)$ such that for $|\lambda| > \lambda_0$, one
has a collection of perturbed potentials $\{
S_\ell=S_\ell(V,\lambda) \}_{\ell=1}^\infty$, $S_\ell \subset
\cS(T^{(\ell)},\delta_\ell)$, $\log T^{(\ell+1)} \asymp \bigl(
T^{(\ell)} \bigr)^\alpha$, \newline $0< \alpha \ll 1$,
$\sum\limits_{\ell=1}^{\infty} \bigl(1-|S_\ell|\bigr) \le
\lambda^{-\beta}$, so that for any potential
$$\widetilde V(x) = V(x) + \sum\limits_{\ell=1}^\infty
 W^{(\ell)}\bigl(\eta^{(\ell)}, \xi^{(\ell)}, \theta^{(\ell)}, \{ R^{(\ell)}_m \} ; x \bigr) $$
where $W^{(\ell)} \in S_\ell$, there exists $\Omega=\Omega(\lambda,\tilde V)$,
$\mes \Omega \le \lambda^{-\beta}$, so that the Lyapunov exponent
$L(\w,E) \ge {1 \over 4} \log \lambda$ for any $\w \in\tor\setminus \Omega,\ E\in \IR$.
Furthermore, the corresponding eigenfunctions are exponentially localized.
\end{theorem}

There are two central technical problems which one has to deal with
in order to establish this theorem.  The first one consists of the
splitting of eigenvalues of the problem (\ref{eq:7.1}) on a finite
interval $[-N,N]$.  The technology for this splitting developed
in~\cite{GS2} for the case of an analytic potential can be modified
for a "generic" smooth potential.

\begin{prop}  Using the notation of Theorem~\ref{thm:chan}, there exist integers
$T'_s$, $\log T'_s \asymp \log T^{(s)}$, such that for any nested
sequence of intervals $\cF_{s,k_s} = [ {k_s \over T_s}, {k_s + 1
\over T_s} )$, and $x\in\tor, \w\in\tor\setminus\Omega$, there is a
sequence of integers $\{ N_s = N_s(x,\w) \}$, with $\log N_{s}
\asymp \log T'_s$, so that
\begin{equation} |E_1-E_2| > \exp (-N_s^{\tau}) \end{equation}
for distinct eigenvalues $E_1,E_2 \in \bigl( \rsp H_{[-N_s,N_s]} (x,\w) \bigr) \cap \cF_{s,k_s}$.
\end{prop}

The second problem is as follows.  The eigenvalues of the problem
(1.1) on a finite interval [1,N] have a parametrization $E_1(x) <
E_2(x) < \ldots < E_N(x),\ x \in \tor$ which are as smooth as the
potential $V(x)$.  This general result is due to the self
adjointness of the the problem (\ref{eq:7.1}) and nondegeneracy of
the the eigenvalues of (\ref{eq:7.1}) restricted on a finite
interval.  The problem is how to evaluate the quantity
\begin{equation}
|\partial_x E_j| + |\partial_{xx} E_j|
\end{equation}
from below.

This problem was also studied in [GS2];  for analytic potentials,
the problem was solved using discriminants of polynomials and
Sard-type arguments.  This method has no modification for smooth
potentials.  This is the very problem for which the variations of
the potential were introduced.  The most basic idea of the method of
\cite{Cha} is as follows.

``Typical'' $C^3$  functions $F(x)$ are Morse functions, i.e., the
quantity
\begin{equation}
|\partial_x F| + |\partial_{xx} F|
\end{equation}
has a good lower bound, gauged according to the size of $F$. On the
other hand, there is a basic relation between $\partial_x E_j$ and
the potential $V(x)$:
\begin{equation}\label{eq:1.99}
\partial_x E_j = \sum\limits_{k=1}^{N} V'(x+k\w) |\varphi_j(x)(k)|^2
\end{equation}
where $\varphi_j(x)(.)$ is a normalized eigenfunction of
(\ref{eq:7.1}) on the interval $[1,N]$ corresponding to $E_j(x)$.
The relation (\ref{eq:1.99}) enables one to express the
``genericity'' of the potential $V$ in terms of a lower bound,
provided $\varphi_j(x)(.)$ is exponentially localized.  Ultimately,
the bad cases can be eliminated by varying the frequencies $\w$. The
Sard-type arguments allow one to show that the total measure of
those $\w$ for which there is no response in (\ref{eq:1.99}) under
the variations of $V$ is extremely small.



\end{document}